\documentclass[a4paper,12pt]{article}
\pdfoutput=1 

\usepackage{jheppub} 

\usepackage[T1]{fontenc} 

\usepackage{amsmath}
\usepackage{dsfont}
\usepackage{xcolor}

\def\be{\begin{equation}}
\def\ee{\end{equation}}

\def\bea{\begin{eqnarray}}
\def\eea{\end{eqnarray}}

\usepackage{amssymb}

\allowdisplaybreaks

\title{Defects in ${\cal N}=1$ minimal models and RG flows}

\author[a]{Matthias R.\ Gaberdiel,}
\author[b]{and Lasse Merkens}

\affiliation[a]{Institut f\"ur Theoretische Physik, ETH Zurich, \\ CH-8093 Z\"urich, Switzerland}
\affiliation[b]{Institut f\"ur Physik, Humboldt-Universit\"at zu Berlin, \\ 10099 Berlin, Germany}

\emailAdd{gaberdiel@itp.phys.ethz.ch}
\emailAdd{lasse.merkens@hu-berlin.de}

\abstract{Utilising the symmetry constraints of suitable topological defects, the possible RG flows of ${\cal N}=1$ superconformal minimal models are studied. We first employ a coset description that only captures the bosonic subalgebra, and then generalise the discussion to the actual superconformal models.}

\begin{document} 
\maketitle
\flushbottom

\section{Introduction}

Generalised symmetries in (2d) conformal field theories have recently attracted a lot of attention since they give rise to interesting insights into the structure of the theory in question, see e.g.\ \cite{Schafer-Nameki:2023jdn,Shao:2023gho} for recent reviews. The paradigm example of a generalised symmetry is a defect line in a 2d CFT, and these have been studied in detail over the years, see e.g.\ \cite{Oshikawa:1996dj,Petkova:2000ip,Fuchs:2002cm,Fuchs:2003id,Fuchs:2004dz,Frohlich:2004ef}. These  defect lines define `generalised' symmetries in that their action on the space of states need not be group-like. Furthermore, the defect lines map local operators to in general (non-local) disorder operators, and thus have a more intricate structure. From the 2d CFT perspective, defect lines have been used to understand non-trivial modular invariants, see e.g.\ \cite{Frohlich:2009gb}. They have also shed light on RG flows, see \cite{Gaiotto:2012np,Chang:2018iay}, and have had interesting applications, for example to 2d YM, see e.g.\ \cite{Cordova:2024goh}. 

More recently, it was proposed in \cite{Nakayama:2024msv} that one may use the topological defects of the Virasoro minimal models in order to constrain the possible RG flows, and this was further studied in \cite{Ambrosino:2025yug};  some of the proposed flows were also tested by other means in  \cite{Katsevich:2024jgq,Katsevich:2024sov,Delouche:2024yuo,Giombi:2024zrt}. In particular, the idea of \cite{Nakayama:2024msv} was to study the subalgebra of those topological defects that commute with the perturbation and hence should be preserved along the RG flow. In particular, this subalgebra must therefore also be present in the IR theory, and this imposes non-trivial constraints about the possible IR fixed-point. 
\smallskip

It is the aim of this paper to generalise the analysis of \cite{Nakayama:2024msv} to the ${\cal N}=1$ superconformal minimal models \cite{Friedan:1984rv,Eichenherr:1985cx}. While much of the analysis is quite parallel to what was done in \cite{Nakayama:2024msv}, there are two subtleties that make this generalisation interesting. First of all, while the ${\cal N}=1$ superconformal minimal models have a coset description just like the bosonic minimal models \cite{Goddard:1986ee,Mathieu:1990dy}, strictly speaking, the relevant coset algebra actually only describes the bosonic subalgebra of the ${\cal N}=1$ superconformal algebra. This coset perspective therefore allows one to treat these models by the usual bosonic methods\footnote{In particular, it allows one to treat both conventional symmetries and R-parity symmetries on an equal footing. The interplay between duality defects and the chiral fermion number symmetry has been analysed for example in \cite{
Thorngren:2018bhj,
Ji:2019ugf,Lin:2019hks,Hsieh:2020uwb},
see  also \cite{Yuji,Seiberg:2023cdc} for reviews.} --- in particular, one sidesteps in this manner some of the subtleties concerning the structure of the Verlinde algebra for superconformal field theories, see \cite{Eholzer:1993ek} --- but the RG constraints one imposes in this manner are slightly weaker in that the full superconformal symmetry is not imposed. 
We shall nevertheless first follow this route since the generalisation to the full superconformal setup turns out to be rather straightforward.  The other subtlety is a consequence of the fact that, unlike the bosonic case, some ${\cal N}=1$ superconformal cosets actually have a fixed-point, which needs to be resolved carefully, e.g.\ following the general prescription given in \cite{Fuchs:1995tq}. 

The rest of the analysis, however, proceeds similarly to what was done in \cite{Nakayama:2024msv}, and the resulting RG flows have  a similar structure  as in the bosonic case. Among other things, they reproduce the RG flows that had been known before, see e.g.\ \cite{Poghossian:1987ngr,Kastor:1988ef,Poghosyan:2014jia}.
\medskip

The paper is organised as follows. In Section~\ref{sec:N=1coset}, we fix our conventions and describe the relevant coset construction. We explain in detail how the fixed-points need to be resolved, and give an explicit formula for the $S$-matrix and the fusion rules, see Section~\ref{sec:2.3}. We also discuss the symmetries of these models, as well as the relevant deformations. In Section~\ref{sec:defects} we construct the defects for the diagonal coset modular invariant; in terms of the ${\cal N}=1$ superconformal symmetry, this theory has a type 0 GSO-projection (and hence is not really superconformal). We also identify the previous symmetries in terms of topological defect lines, see Section~\ref{sec:3.1}. Section~\ref{sec:RGflows} analyses the RG flows of this bosonic coset theory, using the same ideas as in \cite{Nakayama:2024msv}, and the resulting structure is described by eq.~(\ref{mainresult}). In Section~\ref{Supersymmetric defects} we generalise the discussion to the actual superconformal theory which can be described in terms of a non-diagonal modular invariant of the bosonic coset, see eq.~(\ref{susytheory}). Since the modular invariant is of extension type, the analysis generalises directly from the bosonic case, and we   find the RG flows of eq.~(\ref{susyRG}). Finally, Section~\ref{sec:concl} contains our conclusions, and there are two appendices in which we discuss in some detail the fractional level $\mathfrak{su}(2)$ theories that appear for the non-unitary minimal models,  see Appendix~\ref{su2frac}, as well as some rather technical arguments concerning the analysis of the defects, see Appendix~\ref{app:defect}.

\section{${\cal N}=1$ minimal models and their coset description}\label{sec:N=1coset}

\subsection{The ${\cal N}=1$ minimal models}
The ${\cal N}=1$ superconformal algebra consists of the modes of the energy momentum tensor $T(z)$ as well as those of the supercurrent $G(z)$. The corresponding generators satisfy the (anti-)commutation relations 
\begin{align}
[L_m, L_n] &= (m-n)L_{m+n} + \frac{c}{12}(m^3-m)\delta_{m,-n}\ ,  \\
[L_m, G_r] &= \bigl(\tfrac{m}{2} - r\bigr)G_{m+r}\ , \\
\{G_r, G_s\} &= 2L_{r+s} + \frac{c}{3}\bigl(r^2 - \tfrac{1}{4}\bigr)\delta_{r,-s}\ ,  \label{GG}
\end{align}
where $c$ is the central charge, and $r,s\in\mathbb{Z} + \frac{1}{2}$ in the NS sector, while $r,s\in\mathbb{Z}$ in the R sector. 

Just as for the bosonic Virasoro minimal modes, also the ${\cal N}=1$ algebra possesses minimal models. They arise for \cite{Friedan:1984rv,Eichenherr:1985cx}
\begin{align}
c_{p,q} &= \frac{3}{2}\Bigl(1 - \frac{2(p - q)^2}{pq}\Bigr)\ , \label{central charge N=1 min mod} \\
h_{r,s} &=
\begin{cases}
\frac{(qr - ps)^2 - (q - p)^2}{8pq}\ , \qquad & r - s \in 2\mathbb{Z} \quad \quad\ \ (\text{NS sector})\ , \\
\frac{(qr - ps)^2 - (q - p)^2}{8pq} + \frac{1}{16}\ , \qquad & r - s \in 2\mathbb{Z} + 1 \quad (\text{R sector}) \ , 
\end{cases}
\label{susy conformal dimensions}
\end{align}
where $p$ and $q$ are two positive integers, satisfying 
\begin{align}
 p,q\geq 2 \ ,  \qquad q - p \in 2\mathbb{Z}\ , \qquad \text{gcd}\left(p, \frac{|q - p|}{2}\right) = 1\ ,\label{choice p q}
\end{align}
while $(r,s)$ are two integers, taking the values $1\leq r \leq p-1$, $1\leq s \leq q-1$. In the following we shall denote the second entry in the `gcd' by 
\be\label{udef}
u = \frac{|q - p|}{2} \ . 
\ee
The gcd condition then implies that if $p$ and $q$ are even, $u$ must necessarily be odd. 

We should mention that we did not assume that $p<q$, as is often done in the literature. Furthermore,  $\phi_{(1,1)}$ is the identity with $h_{1,1}=0$, and the description in terms of $(r,s)$ is two-fold degenerate since\footnote{In the literature, the roles of $(r,s)$ are sometimes interchanged.}
\be\label{hrsiden}
h_{r,s} = h_{p-r,q-s} \ . 
\ee
Note that, if $p$ and $q$ are even, there is a fixed-point under this identification corresponding to $r=\frac{p}{2}$ and $s=\frac{q}{2}$; the associated highest weight state lies in the Ramond sector (since $\frac{q-p}{2}$ must be odd on account of the last condition in (\ref{choice p q})), and it has conformal dimension equal to 
\be\label{cp}
h_{r=\frac{p}{2}, s = \frac{q}{2}} = \frac{c_{p,q}}{24} \ .
\ee
This state is therefore a `chiral primary', i.e.\ $G_0^2=0$ on this state as follows from (\ref{GG}). We should mention that the definition of $G_0$ on any highest weight state in the Ramond sector $\phi_{\rm R}$ is a bit delicate since $G_0^\dagger = G_0$. We shall work with the convention, see e.g.\ \cite{Watts:1993sj},  that  
\be\label{lambdadef}
G_0\, \phi_{\rm R} = \rho \, \phi_{\rm R} \ ,
\ee
where $\rho$ then determines the conformal dimension via  $h=\rho^2 + \frac{c}{24}$, as follows from \eqref{GG}. For the chiral primary state, $r=\frac{p}{2}, s=\frac{q}{2}$ we then have $\rho=0$. 

Finally, we mention that the \textit{unitary} minimal models arise for $(p, q) = (m, m + 2)$ with $m \geq 2$. Their modular invariants have been classified in \cite{Cappelli:1986ed}.

\subsection{Coset description}

As for the Virasoro minimal models, the ${\cal N}=1$ minimal models can also be described in terms of a diagonal ${\mathfrak{su}}(2)$ coset \cite{Goddard:1986ee,Mathieu:1990dy}, see also \cite{Kastor:1987dk,DiFrancesco:1988xz} 
\begin{align}
    \frac{{\mathfrak{su}}(2)_k\oplus {\mathfrak{su}}(2)_2}{{\mathfrak{su}}(2)_{k+2}} \ , \label{N=1 minimal model cosets}
\end{align}
whose central charge equals 
\be
c_k = \frac{3}{2} \Bigl( 1 - \frac{8}{(k+2) (k+4)} \Bigr)  = c_{k+2,k+4} \ , 
\ee
and hence agrees with the unitary ${\cal N}=1$ minimal model with $m=k+2$. Just as for the non-unitary Virasoro minimal models, the non-unitary ${\cal N}=1$ minimal models also have a coset description where now the level $k$ equals 
\be\label{kdef}
k=\frac{2\max(p,q)}{|q-p|}-2 \ ,  \qquad  \hbox{i.e.} \quad k = \frac{2\min(p,q)}{|q-p|} \ ,
\ee
which is fractional if $|q-p|\neq 2$. We review the salient features of the fractional level $\mathfrak{su}(2)$ theories and their representations in some detail in Appendix~\ref{su2frac}.

For the following it is important to note that the coset only describes the bosonic subalgebra of the ${\cal N}=1$ algebra, and in particular, the NS sector representations of the ${\cal N}=1$ algebra are direct sums of two bosonic coset representations. To see this we need to recall a few facts about the above coset representations. In order to treat both the unitary (for which $k$ is an integer), and the non-unitary case (for which $k$ is fractional) uniformly, it is convenient to write the formulae in terms of the variables 
\be
k^I \equiv p-2 \ , \qquad \hbox{and} \qquad (k+2)^I \equiv q-2 \ , 
\ee
where $k^I$ is defined in Appendix~\ref{su2frac}, and $k^I = k$ for integer $k$, whereas for fractional $k$, $k^I$ is still an integer, see eq.~(\ref{k0kF}). Using the results of Appendix~\ref{su2frac} and in particular eq.~(\ref{A.5}), the representations of the coset theory can be labelled by the triplets 
\be
(\lambda,\mu;\nu) \ , \qquad \lambda\in P^+(\mathfrak{su}(2)_{p-2}) \ , \ \ \mu\in P^+(\mathfrak{su}(2)_{2})  \ , \ \ \nu\in P^+(\mathfrak{su}(2)_{q-2}) \ , 
\ee
where $P^+(\mathfrak{su}(2)_{\hat{k}})$ denotes the integrable weights of $\mathfrak{su}(2)_{\hat{k}}$. (Explicitly, $P^+(\mathfrak{su}(2)_{\hat{k}})$ consists of the weights $\lambda=0,\frac{1}{2},\ldots,\frac{\hat{k}}{2}$.) The triplets must satisfy the selection rule, 
\begin{align}
    \lambda+\mu-\nu\in\mathbb{Z}\ ,\label{selection rule}
\end{align}
and since $\mu\in P^+(\mathfrak{su}(2)_{2})$, the $\mu$ parameter takes the values $\mu\in\{0,\frac{1}{2},1\}$. The conformal dimensions of the coset representations equal 
\begin{align}
        h_{(\lambda,\mu;\nu)}
        &= \lambda(\lambda+1)\frac{|q-p|}{2p}-\nu(\nu+1)\frac{|q-p|}{2q}+n+
        \begin{cases}
            0 & \mu=0\\
            \frac{3}{16} & \mu=\frac{1}{2}\\
            \frac{1}{2} & \mu=1\ ,
        \end{cases}
    \label{nonunitary coset conformal dimesions}
\end{align}
where $n\in\mathbb{N}$ denotes the lowest  level above the ground states in ${\cal H}_\lambda \otimes {\cal H}_\mu$ at which ${\cal H}_\nu$ appears. (Here ${\cal H}_\lambda$ is the affine representation corresponding to $\lambda$.) In terms of the minimal model parameters, see eq.~(\ref{susy conformal dimensions}), the identifications are  
\begin{align}\label{translation}
    r=2\lambda+1 && s=2\nu+1\ .
\end{align}
It follows from the selection rule \eqref{selection rule} that for $\lambda-\nu$ integer, $\mu$ has to be either $0$ or $1$, while for $\lambda-\nu$ half integer we have $\mu=\frac{1}{2}$. Thus $\mu=0,1$ describes the two bosonic subrepresentations of the NS-sector representation, and one of the representations $(\lambda,0;\nu)$ and $(\lambda,1;\nu)$ contains the superconformal primary state, while the other one is its lowest $G$-descendant (which is also primary with respect to the bosonic subalgebra).\footnote{Generically, the lowest $G$-descendant is $G_{-\frac{1}{2}}$ acting on the primary state; the only exception is the vacuum representation for which $G_{-\frac{1}{2}} |0\rangle_{\rm NS}$ is null, and the ground state of the other representation is $G_{-\frac{3}{2}} |0\rangle_{\rm NS}$.}  This fits together with the fact that their conformal dimensions differ by a half-integer, as is clear from (\ref{nonunitary coset conformal dimesions}). In fact, by comparing conformal dimensions, 
see eqs.~\eqref{susy conformal dimensions} and \eqref{nonunitary coset conformal dimesions}, we find that the superconformal primary state in the NS-sector is the one for which $\mu$ takes the value (recall that in the NS-sector $(\lambda-\nu)\in \mathbb{Z}$)
\begin{align}
    \mu&=\begin{cases}
        0 \;\;\; \ \ \text{for }  (\lambda-\nu) \text{ even} \\
        1 \;\;\; \ \ \text{for } (\lambda-\nu) \text{ odd} 
    \end{cases}=\begin{cases}
        0 \;\;\; \ \ \text{for } \frac{(r-s)}{2}  \text{ even}  \\
        1 \;\;\; \ \ \text{for } \frac{(r-s)}{2}  \text{ odd.}
    \end{cases}\label{mu value for susy primary}
\end{align}
We also find that $n$ in \eqref{nonunitary coset conformal dimesions} equals
\be
n =  \frac{(\lambda-\nu)^2}{2}-\frac{\mu}{2} \in \mathbb{N}_0 \qquad \hbox{NS sector, i.e.\ $(\lambda-\nu)\in\mathbb{Z}$.}
\ee
On the other hand, $\mu=\frac{1}{2}$ describes the R-sector representations, for which $n$ takes the form 
\be
n = \frac{(\lambda-\nu)^2-\frac{1}{4}}{2} \in \mathbb{N}_0 \qquad \hbox{R sector, i.e.\ $(\lambda-\nu)\in\mathbb{Z}+\frac{1}{2}$.}
\ee
As long as $\rho \neq 0$, see eq.~(\ref{lambdadef}), any irreducible ${\cal N}=1$ R-sector representation is also irreducible with respect to the bosonic (coset) subalgebra since any fermionic descendant of $\phi_{\rm R}$ can be written as a bosonic descendant of $G_0 \, \phi_{\rm R} = \rho \,  \phi_{\rm R}$. The situation for the chiral primaries (for which $\rho=0$) is more subtle, and we will come back to it momentarily.

Actually not all triplets $(\lambda,\mu;\nu)$ are inequivalent, but we have the field identifications 
\begin{align}
    (\lambda,\mu;\nu)\sim \mathcal{T} (\lambda,\mu;\nu)\equiv (\tau_{k^I}\,\lambda,\tau_2^u\,\mu;\tau_{(k+2)^I}\,\nu)=\bigl(\tfrac{k^I}{2}-\lambda,\tau_2^u\,\mu,\tfrac{(k+2)^I}{2}-\nu\bigr)\ ,\label{field identification}
\end{align}
where $\tau_k$  is the outer automorphism of $\mathfrak{su}(2)_k$ that maps the representation $\lambda \mapsto \frac{k}{2} - \lambda$, and $u = \frac{|q-p|}{2}$. Note that this is just the coset analogue of the identification of eq.~(\ref{hrsiden}). We shall use the convention that the equivalence class defined by the above identification will be denoted by $[(\lambda,\mu;\nu)]$.

The representations $[(\lambda,\mu;\nu)]$ are all irreducible, except for the representation $(\frac{p-2}{4},\frac{1}{2};\frac{q-2}{4})$ that is a fixed-point under the field identification (\ref{field identification}).\footnote{The general procedure of decomposing the fixed-point representation is usually referred to as the `fixed-point resolution' of the coset, and it has implications for the characters, the modular $S$-matrix, and the fusion rules, see below; a general discussion of it can be found in \cite{Fuchs:1995tq}.} To understand this we first note that the corresponding  
${\cal N}=1$ superconformal R-sector representation --- this is described by $(r=\frac{p}{2}, s = \frac{q}{2})$, see eq.~(\ref{cp}) ---  is irreducible. However, it is not irreducible with respect to the bosonic subalgebra (that is described by the coset), but rather decomposes into two subspaces that we denote as 
\be\label{fixedpointreso}
(r=\tfrac{p}{2}, s = \tfrac{q}{2}) \cong (\tfrac{p-2}{4},\tfrac{1}{2};\tfrac{q-2}{4})_+ \oplus (\tfrac{p-2}{4},\tfrac{1}{2};\tfrac{q-2}{4})_- \ . 
\ee
These two subspaces are the eigenspaces of $(-1)^F = \pm 1$, where $F$ is the fermion number, see eq.~(\ref{Ferm})  below. Generically, i.e.\ if $\rho\neq 0$, this operator does not make sense for a R sector representation since the highest weight state $\phi_{\rm R}$ does not have a well-defined fermion number as follows from (\ref{lambdadef}). However, when $\rho=0$, as is the case for $(r=\tfrac{p}{2}, s = \tfrac{q}{2})$, one can define $(-1)^F \phi_{\rm R} = + \phi_{\rm R}$, say, and then the fermion number of all descendants is unambiguously defined. Since $(-1)^F$ commutes with the bosonic coset algebra, its eigenspaces therefore define (irreducible) subrepresentations with respect to this bosonic subalgebra, and this gives rise to (\ref{fixedpointreso}).

\subsection{The $S$-matrix and the fusion rules}\label{sec:2.3}

The $S$-matrix of the coset characters takes the form\footnote{To avoid clutter we shall label the matrix elements of the $S$-matrix in terms of the individual triplets $(\lambda,\mu;\nu)$, rather than their equivalence classes. In fact, the formula is the same for the fields that are related by the field identification.}
\begin{align}
    \begin{split}
        \mathcal{S}_{(\lambda,\mu;\nu),(\lambda^\prime,\mu^\prime;\nu^\prime)}=&\sqrt{\frac{8}{pq}}\sin\left(\pi(2\lambda+1)(2\lambda^\prime+1)\frac{|q-p|}{2p}\right)\\
        &\times\sin\left(\frac{\pi(2\mu+1)(2\mu^\prime+1)}{4}\right)\sin\left(\pi(2\nu+1)(2\nu^\prime+1)\frac{|q-p|}{2q}\right)\ , 
  \end{split}\label{coset S matrix}
\end{align}
provided that neither $(\lambda,\mu;\nu)$ nor $(\lambda^\prime,\mu^\prime;\nu^\prime)$ are equal to $(\frac{p-2}{4},\frac{1}{2};\frac{q-2}{4})_\pm$. (Obviously, these special fields only exist provided that $p$ and $q$ are even.) The special matrix elements involving these fixed-point-resolved fields are 
\begin{align}
         \mathcal{S}_{(\lambda,\mu;\nu),(\frac{p-2}{4},\frac{1}{2};\frac{q-2}{4})_\pm}=&\, \mathcal{S}_{(\frac{p-2}{4},\frac{1}{2};\frac{q-2}{4})_\pm,[(\lambda,\mu;\nu)]} \nonumber \\        =& \, 
        \sqrt{\frac{2}{pq}}\sin\left(\pi(2\lambda+1)\frac{|q-p|}{4}\right) \nonumber \\ &
         \times\sin\left(\frac{\pi(2\mu+1)}{2}\right)\sin\left(\pi(2\nu+1)\frac{|q-p|}{4}\right) \ , \label{2.22}\\
        \mathcal{S}_{(\frac{p-2}{4},\frac{1}{2};\frac{q-2}{4})_\pm,(\frac{p-2}{4},\frac{1}{2};\frac{q-2}{4})_{\pm^\prime}}= &
        \begin{cases}
            \frac{1}{2} \quad \text{if }\pm=\pm^\prime\\
            -\frac{1}{2} \quad \text{if }\pm\neq\pm^\prime \ ,
        \end{cases}
\label{coset S matrixspecial}
\end{align}
where in eq.~(\ref{2.22}) $(\lambda,\mu;\nu)$ is a generic coset field, i.e.\ not equal to the special fixed-point fields.

The structure of the corresponding Verlinde formula depends on whether $p$ and $q$ are even, i.e.\ whether there is a fixed-point, or not. In the simpler case that $p$ and $q$ are odd, there are no fixed-points, and the $S$-matrix factorises into three affine ${\mathfrak{su}}(2)$ $S$-matrices. Then the resulting fusion rules are simply 
\be\label{fusionrules}
[(\lambda_1,\mu_1;\nu_1)] \otimes [(\lambda_2,\mu_2;\nu_2)] = \bigoplus_{(\lambda_3,\mu_3;\nu_3)}{\cal N}_{\lambda_1 \lambda_2}^{[k^I]\,\, \lambda_3} \, {\cal N}_{\mu_1 \mu_2}^{[2]\,\, \mu_3} \, 
{\cal N}_{\nu_1 \nu_2}^{[(k+2)^I]\,\, \nu_3} \, [(\lambda_3,\mu_3;\nu_3)] \ ,
\ee
where ${\cal N}_{\lambda_1 \lambda_2}^{[\hat{k}]\,\, \lambda_3}$ are the usual fusion rules of $\mathfrak{su}(2)_{\hat{k}}$, 
\be
{\cal N}_{\lambda_1 \lambda_2}^{[\hat{k}]\,\, \lambda_3} = \left\{ \begin{array}{cl} 1 & \hbox{if $|\lambda_1- \lambda_2| \leq \lambda_3 \leq \min(\lambda_1+\lambda_2,\hat{k}-\lambda_1 -\lambda_2)$ \& 
$\lambda_3 - \lambda_1 - \lambda_2 \in \mathbb{Z}$} \\
0 & \hbox{otherwise,}
\end{array}
\right.
\ee
and we sum over all the triplets $(\lambda_3,\mu_3;\nu_3)$ without any field identifications. As a consequence, the coset fusion rule coefficients may be equal to $2$, namely if both $(\lambda_3,\mu_3;\nu_3)$ and $\mathcal{T}(\lambda_3,\mu_3;\nu_3)$ appear in the $\mathfrak{su}(2)$ fusion rules.

For $p$ and $q$ even, the answer is more complicated since we have the two fixed-point resolved fields with labels $(\frac{p-2}{4},\frac{1}{2},\frac{q-2}{4})_{\pm}$. If $(\lambda_1,\mu_1;\nu_1)$ and $(\lambda_2,\mu_2;\nu_2)$ are both generic coset fields, i.e.\ not equal to $(\frac{p-2}{4},\frac{1}{2},\frac{q-2}{4})_{\pm}$, then eq.~(\ref{fusionrules}) continues to hold provided we replace on the right-hand side
\be\label{sum}
[(\tfrac{p-2}{4},\tfrac{1}{2},\tfrac{q-2}{4})] \mapsto (\tfrac{p-2}{4},\tfrac{1}{2},\tfrac{q-2}{4})_+ \oplus  (\tfrac{p-2}{4},\tfrac{1}{2},\tfrac{q-2}{4})_- \ , 
\ee
i.e.\ in these fusion rules, the two fixed-point-resolved fields always appear together. By the same token, the fusion rules of the sums (\ref{sum}) are the same as those described in (\ref{fusionrules}), but the fusion rules of the individual fixed-point-resolved fields are more complicated, and we have spelled them out in Appendix~\ref{Fusion rules for even models}.

\subsection{Symmetries}\label{sec:symmetries}

For the following it will be important to understand and describe the symmetries of these coset models.  First of all, it is clear that the fusion rules are invariant under the $\mathbb{Z}_2$ symmetry (`spacetime fermion number') that acts as 
\be\label{2.30}
(-1)^{F_s}:\quad  [(\lambda,\mu;\nu)] \mapsto (-1)^{2\mu}\,  [(\lambda,\mu;\nu)] \ . 
\ee
This simply acts as $+1$ on all NS-sector representations (i.e.\ the representations for which $\mu=0,1$), and as $-1$ on all R-sector representations (i.e.\ the representations for which $\mu=\frac{1}{2}$). On the other hand, the definition of a `chiral fermion number' in the NS-sector for which the states with $\mu=0$ and $\mu=1$ have opposite fermion number --- these states are related to one another by the action of an odd number of supercurrent $G$ modes --- is more delicate because of the field identification rules. Since the fusion of any NS-sector field with itself leads to states with $\lambda,\nu\in\mathbb{Z}$ and $\mu=0$, as follows from (\ref{fusionrules}), the chiral fermion number operator must act as $+1$ on all such states, i.e.\
\be\label{Ferm}
(-1)^F:\   (\lambda,0;\nu) \mapsto (\lambda,0;\nu) \ , \qquad \lambda,\nu\in\mathbb{Z} \ . 
\ee
The further analysis depends on whether $u=\frac{|q-p|}{2}$ is even or odd, so let us discuss the two cases in turn.

\subsubsection{The case $u=\frac{|q-p|}{2}$ odd}\label{sec:2.4.1}

We first want to show that if $p$ and $q$ are even, there is no consistent definition of $(-1)^F$  on the entire spectrum. To see how this goes, we note that $p$ and $q$ even imply that $u$ is odd, see eq.~(\ref{udef}). For $u$ odd, the field identification rules relate $(\lambda,0;\nu) \sim (\tfrac{p-2}{2} - \lambda,1; \tfrac{q-2}{2} - \nu)$, and it follows from eq.~(\ref{Ferm}) that 
\be
(-1)^F:\   (\tfrac{p-2}{2} - \lambda,1; \tfrac{q-2}{2} - \nu) \mapsto  + (\tfrac{p-2}{2} - \lambda,1; \tfrac{q-2}{2} - \nu) \ , \qquad \lambda,\nu\in\mathbb{Z} \ .
\ee
Since $G$ anticommutes with $(-1)^F$ this then implies
\be
(\tfrac{p-2}{2} - \lambda,0; \tfrac{q-2}{2} - \nu) \mapsto  - (\tfrac{p-2}{2} - \lambda,0; \tfrac{q-2}{2} - \nu) \ , \qquad \lambda,\nu\in\mathbb{Z} \ .
\ee
However, if $p$ and $q$ are even, this is now in contradiction to eq.~(\ref{Ferm}). We therefore conclude that, for $p$ and $q$ even, there is no consistent assignment of which of the two states $\mu=0$ and $\mu=1$ should be a `boson' and which a `fermion'. On the other hand, {\bf if $p$ and $q$ are odd and $u$ is odd}, we can define a non-trivial chiral fermion number operator on the NS-sector states by 
\be\label{mF}
(-1)^{F} [(\lambda,\mu;\nu)]= (-1)^{2\lambda+\mu} [(\lambda,\mu;\nu)] \qquad \lambda+ \nu \in \mathbb{Z} \ (\hbox{NS sector})\ .
\ee
It is worth pointing out that, because of (\ref{mu value for susy primary}), the ground state is not always the boson (and the first excited $G$-descendant the fermion). The extension to the R-sector is uniquely fixed since  $[(0,1;0)]$ has $(-1)^F$ eigenvalue $-1$, but maps each R-sector representation to itself; thus we have to set 
 \be\label{2.28}
(-1)^{F} [(\lambda,\mu;\nu)]= 0 \qquad \qquad \lambda+ \nu \in \mathbb{Z}+\tfrac{1}{2} \ (\hbox{R sector})\ . 
\ee

\subsubsection{The case $u=\frac{|q-p|}{2}$ even}\label{sec:2.4.2}

First of all, $u$ even is only possible if $p$ and $q$ are odd, see eq.~(\ref{udef}), so we continue to work with the case that {\bf $p$ and $q$ are odd}. For $u$ even the field identification does not relate the $\mu=0$ to the $\mu=1$ states, and hence we can consistently define 
\be\label{worldsheet fermion number for u even}
(-1)^{F} [(\lambda,\mu;\nu)]=\left\{\begin{array}{cl}
 (-1)^\mu \, [(\lambda,\mu;\nu)] \qquad & \lambda-\nu\in\mathbb{Z} \ \hbox{(NS sector)} \\
 0 \qquad & \lambda-\nu\in\mathbb{Z}+\tfrac{1}{2} \ \hbox{(R sector),} 
 \end{array} \right.
\ee
where the expression for the R-sector representations follows from the same arguments as in (\ref{2.28}).
 
 \subsection{Relevant deformations}\label{sec:relevant}
 
 Since we are interested in RG flows relating ${\cal N}=1$ minimal models to one another, we also need to understand which deformation fields are relevant. It will be convenient to label the minimal model before the RG flow by $(p,q)$, where we write $q=sp + I$ with $s\in\mathbb{N}$ and $0\leq I < p$. We shall mainly be interested in the case where the deforming field $[(l,m;n)]$ sits in a representation for which $l=0$. Then the conformal dimension of $[(0,m;n)]$ is less than $1$  (i.e.\ the field is relevant) as long as $n \leq s =\lceil\frac{q}{p}\rceil$, and bigger than $1$ (i.e.\ irrelevant) for $n>s$. 
 
 The natural deformation field we shall primarily have in mind is $[(0,0;s)]$ if $s$ is odd, and $[(0,1;s)]$ if $s$ is even. This is the least relevant relevant deformation, i.e.\ the relevant field with the largest conformal dimension, as long as $q\leq \frac{3}{2} p$.\footnote{If $q>\frac{3}{2}p$, the least relevant relevant field is actually $[(0,0;s+1)]$ for odd $s$, and $[(0,1;s+1)]$ for even $s$. However, also $[(0,0;s)]$ resp.\ $[(0,1;s)]$ continue to be relevant even for $q>\frac{3}{2}p$, and their conformal dimension is only marginally lower.\label{foot1}} Note that, because of (\ref{mu value for susy primary}), these fields are always $G_{-\frac{1}{2}}$-descendants; in fact, in terms of the minimal model description of eq.~(\ref{central charge N=1 min mod}), they always correspond to $G_{-\frac{1}{2}} \phi_{(1,2s+1)}$, and this ties in naturally with what one would expect in the supersymmetric setting.\footnote{Conversely, for $q>\frac{3}{2}p$, the least relevant relevant fields of Footnote~\ref{foot1} are not $G$-descendants, but the ${\cal N}=1$ primary fields $\phi_{(1,2s+3)}$.}

We should also mention that the known RG flows between unitary ${\cal N}=1$ minimal models $(p,p+2) \rightarrow (p,p-2)$ \cite{Poghossian:1987ngr,Kastor:1988ef,Poghosyan:2014jia} are induced by the deformation $G_{-\frac{1}{2}} \phi_{(1,3)}$; in the above parametrisation this corresponds to the case $s=1$ and $I=2$.

\section{The defects of the bosonic coset theory}\label{sec:defects}

In preparation for the discussion of the RG flows of these ${\cal N}=1$ minimal models, we now want to study their defects. The structure of the defects depends somewhat on which modular invariant one considers, and in this section we shall concentrate on the case of the charge conjugation modular invariant of the coset theory with spectrum\footnote{Note that the coset representations are all self-conjugate, and thus there is no distinction between the `charge conjugation modular invariant' and the `diagonal modular invariant'.}
\be\label{cosettheory}
{\cal H} = \bigoplus_{[(l,m;n)]} {\cal H}_{[(l,m;n)]} \otimes \tilde{{\cal H}}_{[(l,m;n)]} \ , 
\ee
where the sum runs over all the inequivalent coset representations (taking the field identification into account). From the point of view of the ${\cal N}=1$ superconformal symmetry this is the GSO-projected theory where we impose the GSO-projection
\be\label{GSO}
\frac{1}{2} \bigl( 1 + (-1)^{F+\tilde{F}} \bigr) \ .
\ee
We should mention that while $(-1)^F$ cannot, in general, be consistently defined on the chiral representations, see the discussion in Section~\ref{sec:symmetries} above, the combination $(-1)^{F+\tilde{F}}$ can be unambiguously defined by setting it to be $+1$ on the ground states of the diagonal NS-NS sector ${\cal N}=1$ superconformal representations. Furthermore, in the R-R sector we now have two fermionic zero modes, $G_0$ and $\bar{G}_0$, and hence before GSO-projection a $2$-dimensional space of ground states, of which one survives the GSO-projection.

Any rational bosonic conformal field theory with a charge conjugation modular invariant has topological defects that are labelled by the representations of the chiral algebra \cite{Petkova:2000ip}, and we denote these simple topological defects by $\mathcal{L}_{[(\lambda,\mu;\nu)]}$, where $[(\lambda,\mu;\nu)]$ labels the different coset representations. If we denote the by $|\phi_{[(l,m;n)]}\rangle$ the (ground) state in the sector
\be\label{3.2}
|\phi_{[(l,m;n)]}\rangle \in {\cal H}_{[(l,m;n)]} \otimes \tilde{{\cal H}}_{[(l,m;n)]} \ , 
\ee
then the defect acts on these states as 
\begin{align}\label{actionform}
    \mathcal{L}_{[(\lambda,\mu;\nu)]}|\phi_{[(l,m;n)]}\rangle =\frac{{\cal S}_{(\lambda,\mu;\nu)(l,m;n)}}{{\cal S}_{(0,0;0)(l,m;n)}}\, |\phi_{[(l,m;n)]}\rangle \ , 
\end{align}
where ${\cal S}$ denotes the $S$-matrix, see eq.~(\ref{coset S matrix}). Furthermore, these topological defect lines satisfy the same fusion algebra as the chiral fields, i.e.\
\be
 \mathcal{L}_{[(\lambda_1,\mu_1;\nu_1)]} \ \cdot \  \mathcal{L}_{[(\lambda_2,\mu_2;\nu_2)]} = 
 \sum_{[(\lambda_3,\mu_3;\nu_3)]} {\cal N}_{[(\lambda_1,\mu_1;\nu_1)], [(\lambda_2,\mu_2;\nu_2)]}{}^{[(\lambda_3,\mu_3;\nu_3)]} \, \mathcal{L}_{[(\lambda_3,\mu_3;\nu_3)]} \ , 
\ee
where the fusion rules were given in Section~\ref{sec:2.3} above. 

\subsection{Simple defects and their symmetries}\label{sec:3.1}

It is easy to identify the defects that correspond to the symmetries we discussed earlier in Section~\ref{sec:symmetries}. First of all, the defect associated to the vacuum representation of the coset $[(0,0;0)]$ acts as the identity, while it follows directly from the structure of the $S$-matrix of eq.~(\ref{coset S matrix}) that the defect corresponding to $(-1)^{F_s}$, see eq.~(\ref{2.30}), is the one associated to the coset representation $[(0,1;0)]$, 
\begin{align}\label{010act}
    \mathcal{L}_{[(0,1;0)]}|\phi_{[(l,m;n)]}\rangle=
    \begin{cases}
        \;\;|\phi_{[(l,m;n)]}\rangle & \text{if } |\phi_{[(l,m;n)]}\rangle\in\text{NS}\\
        -|\phi_{[(l,m;n)]}\rangle & \text{if } |\phi_{[(l,m;n)]}\rangle\in\text{R}\ .
    \end{cases}
\end{align}
This is indeed a $\mathbb{Z}_2$ symmetry since 
\be
\mathcal{L}_{[(0,1;0)]} \cdot  \mathcal{L}_{[(0,1;0)]} =  \mathcal{L}_{[(0,0;0)]} \ . 
\ee
The fusion rules also imply that 
\begin{align}
  \mathcal{L}_{[(\lambda,1;\nu)]}=\mathcal{L}_{[(0,1;0)]} \cdot \mathcal{L}_{[(\lambda,0;\nu)]}\ ,
\end{align}
from which it follows that $\mathcal{L}_{[(\lambda,0;\nu)]}$ and $\mathcal{L}_{[(\lambda,1;\nu)]}$ agree on NS-NS sector states, but differ by a sign in the R-R sector. Similarly, any defect $  \mathcal{L}_{[(\lambda,\mu;\nu)]}$ with $\mu=0$ or $\mu=1$ satisfies
\be\label{3.8}
\mathcal{L}_{[(\lambda,\mu;\nu)]} \, |\phi_{[(l,0;n)]}\rangle = \mathcal{L}_{[(\lambda,\mu;\nu)]} \, |\phi_{[(l,1;n)]}\rangle \ , \qquad \hbox{if $\mu=0$ or $\mu=1$.}
\ee
In particular, this therefore means that any such defect commutes with the action of the supercharge (in the NS-sector). On the other hand, if $\mu=\frac{1}{2}$, the defect anti-commutes with the action of the supercharge (in the NS-sector), i.e.
\be
\mathcal{L}_{[(\lambda,\frac{1}{2};\nu)]} \, |\phi_{[(l,0;n)]}\rangle = - \mathcal{L}_{[(\lambda,\frac{1}{2};\nu)]} \, |\phi_{[(l,1;n)]}\rangle \ .
\ee

\subsection{The chiral fermion number defect}\label{sec:worldsheetferm}

The analysis for the chiral fermion number operator, $(-1)^F$, is more subtle (but also more interesting). As we have seen in Section~\ref{sec:symmetries} above, we can only define this symmetry for {\bf $p$ and $q$ odd}, and the structure of the action depends on whether $u$ is odd (see Section~\ref{sec:2.4.1}) or even (see Section~\ref{sec:2.4.2}). From the perspective of the defects, however, there is a uniform description of both cases: the relevant defect is 
$\mathcal{L}_{[(\frac{p-2}{2},\frac{1}{2};0)]}$, and it acts as 
\begin{align}
    \mathcal{L}_{[(\frac{p-2}{2},\frac{1}{2};0)]}|\phi_{[(l,m;n)]}\rangle=
    \begin{cases}
        \sqrt{2}(-1)^{\frac{|q-p|}{2}(2l+1)+1+m}|\phi_{[(l,m;n)]}\rangle & \text{if } |\phi_{[(l,m;n)]}\rangle\in \text{NS}\\
        0 & \text{if } |\phi_{[(l,m;n)]}\rangle\in \text{R} \ . 
    \end{cases}
\end{align}
The relation to the previously defined chiral fermion number operator is therefore 
\be
\mathcal{L}_{[(\frac{p-2}{2},\frac{1}{2};0)]}=\mathcal{L}_{[(0,\frac{1}{2};\frac{q-2}{2})]}= (-1)^{u+1}\, \sqrt{2}(-1)^{F}\ .
\ee
The factor of $\sqrt{2}$ is necessary for these defects to satisfy the Tambara-Yamagami fusion rule
\begin{align}\label{3.12}
    \mathcal{L}_{[(\frac{p-2}{2},\frac{1}{2};0)]} \cdot \mathcal{L}_{[(\frac{p-2}{2},\frac{1}{2};0)])}={\bf 1} +\mathcal{L}_{[(0,1;0)]} \ , 
\end{align}
where ${\bf 1} = \mathcal{L}_{[(0,0;0)]}$ is the identity defect, while $\mathcal{L}_{[(0,1;0)]} = (-1)^{F_s}$. Thus, for $p$ and $q$ both odd, the chiral fermion number operator (suitably rescaled) squares to the sum of two operators, one of which is the spacetime fermion number operator $(-1)^{F_s}$.
\medskip

On the other hand, for {\bf $p$ and $q$ even} the natural analogue of $(-1)^F$ is the defect $\mathcal{L}_{[(\frac{p-2}{2},0;0)]}$ that now commutes with the supercharge, see eq.~(\ref{3.8}), and hence cannot be thought of as $(-1)^{F}$. (This also reflects what we saw in Section~\ref{sec:2.4.1}.) Together with the defect line $\mathcal{L}_{[(0,1;0)]}$ it satisfies the relations of $\mathbb{Z}_2\times \mathbb{Z}_2$, 
\begin{align}
\mathcal{L}_{[(0,1;0)]} \cdot \mathcal{L}_{[(0,1;0)]} & = {\bf 1} \label{3.13} \\ 
\mathcal{L}_{[(\frac{p-2}{2},0;0)]} \cdot \mathcal{L}_{[(\frac{p-2}{2},0;0)]} & = {\bf 1} \\
\mathcal{L}_{[(0,1;0)]} \cdot \mathcal{L}_{[(\frac{p-2}{2},0;0)]} & = \mathcal{L}_{[(\frac{p-2}{2},0;0)]} \cdot \mathcal{L}_{[(0,1;0)]} = \mathcal{L}_{[(\frac{p-2}{2},1;0)]} \ . \label{3.15}
\end{align}
We should mention that for $q=2$ (or equivalently $p=2$), the two defects $\mathcal{L}_{[(0,1;0)]}$ and $\mathcal{L}_{[(\frac{p-2}{2},0;0)]}$ actually agree because of the field identification. These theories therefore only possess a $\mathbb{Z}_2$ symmetry instead of the $\mathbb{Z}_2\times\mathbb{Z}_2$ symmetry described above.\footnote{This was, for example, already noticed for the ${\cal N}=1$ $(2,8)$ super-Lee-Yang model  in \cite{Klebanov_2023,Nakayama_2021}, which  is actually equivalent to the $(3,8)$ Virasoro minimal model, and thus only possesses a $\mathbb{Z}_2$ symmetry.}
\smallskip

Incidentally, the structural difference between $p$ even and odd also has a natural interpretation in the Landau-Ginzburg (LG) description of the unitary ${\cal N}=1$ mi\-ni\-mal models. Recall that the LG theory with superpotential $W=\Phi^p$ flows in the IR to the unitary ${\cal N}=1$ minimal model $(p,p+2)$. When $p$ is even, $\Phi\mapsto-\Phi$ describes an ordinary $\mathbb{Z}_2$ symmetry, which corresponds to $\mathcal{L}_{[(\frac{p-2}{2},0;0)]}$ in our description. On the other hand, if $p$ is odd, $\Phi\mapsto-\Phi$ is still a symmetry, but it acts as an R-parity since it anti-commutes with the supercurrent. (It realises the defect $\mathcal{L}_{[(\frac{p-2}{2},\frac{1}{2};0)]}$ in our language.) This R-parity possesses a 't~Hooft anomaly, and this is reflected in the fact that, in the bosonic description, it is non-invertible, see eq.~(\ref{3.12}) \cite{Thorngren:2018bhj,Ji:2019ugf,Lin:2019hks,Hsieh:2020uwb};  for a discussion of this in the context of the $(3,5)$ ${\cal N}=1$ minimal model see also \cite{Kikuchi:2022jbl,Nakayama_2023}.


\section{RG flows of ${\cal N}=1$ minimal models}\label{sec:RGflows}

We are now ready to analyse the RG flows of the ${\cal N}=1$ minimal models, using similar techniques as employed in \cite{Nakayama:2024msv}, who studied the corresponding case for the Virasoro minimal models, see also \cite{Ambrosino:2025yug}. The basic idea is to constrain the possible RG flows by using the topological defects that are preserved by the flow.

\subsection{Renormalisation group invariants}

We are interested in perturbing the above coset theory by a relevant operator $\phi$. The idea of the analysis is to identify the subset of defects $\mathcal{L}$ that commutes with the defect in the sense of 
\begin{align}\label{preservation}
    \mathcal{L}\, \phi (z)\, |\psi\rangle=\phi(z)\, \mathcal{L}  \, |\psi\rangle\ ,
\end{align}
where $|\psi\rangle$ is an arbitrary state in the space of states of the theory. In particular, then the topological defect line remains a topological defect line along the whole RG flow, and in particular, also defines a defect line of the IR theory. Since the preserved defects form a fusion ring, a necessary condition for a specific RG flow to exist is that both the UV and the IR theory contain an isomorphic ring of preserved topological defect operators. 

We will study these constraints in detail, but in order to exclude possible flows already a simpler condition is often sufficient. Given that the topological defect lines ${\cal L}_{[(\lambda,\mu;\nu)]}$ satisfy the fusion rules of the minimal model, their quantum dimension, $d_{[(\lambda,\mu;\nu)]}$, defined via 
\begin{align}
    \mathcal{L}_{[(\lambda,\mu;\nu)]} \, |0\rangle= d_{[(\lambda,\mu;\nu)]}\, |0\rangle 
\end{align}
must also do so. Indeed, given the form of eq.~(\ref{actionform}), the quantum dimension of the defect line associated to $[(\lambda,\mu;\nu)]$ is simply\footnote{In the non-unitary case, one may also consider replacing the first vacuum representation in the denominator of eq.~(\ref{qdim}) by the representation of smallest conformal weight, see e.g.\  \cite{Fukusumi:2025xrj} for a recent discussion. However, we have found that the constraints arising from the quantum dimension that is defined relative to the actual vacuum are strongest. This is probably related to the fact that the vacuum is not the state of lowest conformal dimension, and its conservation is therefore non-trivial. We thank Yoshiki Fukusumi for a comment about this point.} 
\be\label{qdim}
d_{[(\lambda,\mu;\nu)]} = \frac{{\cal S}_{(\lambda,\mu;\nu)(0,0;0)}}{{\cal S}_{(0,0;0)(0,0;0)}} \ , 
\ee
and it follows directly from the Verlinde formula that these ratios form a one-dimensional representation of the fusion algebra,
\begin{align}
    d_{[(\lambda_1,\mu_1;\nu_1)]} \cdot d_{[(\lambda_2,\mu_2;\nu_2)]}  = 
    \sum_{[(\lambda_3,\mu_3;\nu_3)]}  {\cal N}_{[(\lambda_1,\mu_1;\nu_1)], [(\lambda_2,\mu_2;\nu_2)]}{}^{[(\lambda_3,\mu_3;\nu_3)]} \,  d_{[(\lambda_3,\mu_3;\nu_3)]} \ . \label{qdim fusion algebra}
\end{align}
The one-dimensional representations of the fusion algebra are discrete, and therefore the quantum dimensions of the preserved defect lines must be the same along the flow, and therefore, in particular, agree between the UV and the IR CFT. 

\subsubsection{The spin content of the twisted Hilbert space}

Another convenient invariant we shall consider, again following \cite{Nakayama:2024msv}, is the 
spin content of the twisted Hilbert space $\mathcal{H}_\mathcal{L}$. As was already explained in \cite{Petkova:2000ip}, a topological defect line can be used to twist a given theory, and this leads to a twisted Hilbert space $\mathcal{H}_\mathcal{L}$ that can be decomposed as 
\be
{\cal H}_{\cal L} = \bigoplus_{i,j} M_{i,j}({\cal L}) \, {\cal H}_i \otimes \tilde{\cal H}_j \ , 
\ee
where the sum over $i$ and $j$ runs over the irreducible representations of the chiral algebra. Since this Hilbert space does not describe local operators, the left- and right-conformal dimensions do not need to differ by an integer, but instead $h_i - h_j$ may have a non-trivial fractional part. We call this fractional part the `spin', and the spin-content of the twisted Hilbert space is then the set
\be
{\rm Spin}_{\cal L} = \left\{ h_i - h_j  \, \, \hbox{mod} \, \mathbb{Z}: \, M_{i,j}({\cal L}) \neq 0 \right\} \ . 
\ee
It has been argued in \cite{Chang:2018iay,Kikuchi:2022jbl} that the defect Hilbert space spin content of a preserved topological defect line is also an RG invariant.\footnote{It was argued in \cite{Kikuchi:2022jbl} that the spin content of the IR theory must only be a subset of that of the UV theory. However, for our examples, the spin content of the IR and the UV theory will turn out to agree.}

\subsection{Flows of the form $\mathcal{SM}(p,q)\to\mathcal{SM}(p,q^\prime)$}

With these preparations we can now study the possible flows of the above coset theories. To streamline notation we shall denote the diagonal coset theory corresponding to $(p,q)$ by $\mathcal{SM}(p,q)$. Suppose we perturb $\mathcal{SM}(p,q)$ by the relevant field $\phi_{[(l,m;n)]}$ in the space of states of $\mathcal{SM}(p,q)$, see eq.~(\ref{3.2}). We are interested in flows whose endpoint is yet another conformal field theory (rather than a massive theory), and we denote this schematically by 
\be
\mathcal{SM}(p,q)\to\mathcal{SM}(p^\prime,q^\prime) \ . 
\ee
For suitable choices of $\phi_{[(l,m;n)]}$, we want to use the above constraints to make a prediction for what $(p',q')$ should be. 

The first observation is that for a {\bf generic} perturbation $\phi_{[(l,m;n)]}$, only very few defects are preserved, and hence this method will not be very powerful. Obviously, $\mathcal{L}_{[(0,0;0)]}={\bf 1}$ is always preserved, and $\mathcal{L}_{[(0,1;0)]}$ is preserved provided that $m=0$ or $m=1$. If $p$ and $q$ are odd, then $\mathcal{L}_{[(\frac{p-2}{2},\frac{1}{2};0)]}$ is preserved if $\phi_{[(l,m;n)]}$ is a boson, i.e.\ if it has eigenvalue $+1$ with respect to (\ref{mF}). On the other hand, if $p$ and $q$ are even (and hence $u$ is odd), $\mathcal{L}_{[(\frac{p-2}{2},0;0)]}$ is only preserved if $l\in\mathbb{Z}$. All other lines are generically not preserved.

The invariants arising from these defects give rather few constraints. First of all, every theory possesses the lines $\mathcal{L}_{[(0,0;0)]}={\bf 1}$ and $\mathcal{L}_{[(0,1;0)]}$, and hence their existence does not impose any constraints on $(p',q')$. The corresponding defects satisfy very simple fusion algebras, see eq.~(\ref{3.12}) and eqs.~(\ref{3.13}) -- (\ref{3.15}) above, and this characterises these defects effectively uniquely. This is also reflected in the fact that their quantum dimension are very simple, namely
\be\label{4.8}
d_{[(0,0;0)]} = d_{[(0,1;0)]} = 1 \ , \qquad  d_{[(\frac{p-2}{2},0;0)]} = (-1)^{u+1}  \ , \quad d_{[(\frac{p-2}{2},\frac{1}{2};0)]} = (-1)^{u+1} \, \sqrt{2} \ .
\ee
Using these constraints we can therefore deduce that 
\begin{list}{\arabic{enumi}.}{\usecounter{enumi}}
\item If $p$ and $q$ are odd and $\phi_{[(l,m;n)]}$ is a boson, then the existence of $\mathcal{L}_{[(\frac{p-2}{2},\frac{1}{2};0)]}$ implies that also $p'$ and $q'$ are odd, and that $u'-u$ is even. 
\item  If $p$ and $q$ are even, and $l\in\mathbb{Z}$, then the existence of $\mathcal{L}_{[(\frac{p-2}{2},0;0)]}$ implies that also $p'$ and $q'$ are even. 
\end{list}

\subsubsection{Deformations by $\phi_{[(0,m;n)]}$}\label{sec:deformations}

In order to get interesting constraints we therefore need to consider more special perturbations, and an interesting class is described by the deformations by  $\phi_{[(0,m;n)]}$; in fact, this accounts for the  least relevant relevant deformations, see the discussion in Section~\ref{sec:relevant}, and in particular, includes the integrable deformations that have been studied before in \cite{Poghossian:1987ngr,Kastor:1988ef,Poghosyan:2014jia}. 

Depending on the value of $m$, the perturbation by $\phi_{[(0,m;n)]}$ preserves the following topological defect lines
\begin{align}
m=\tfrac{1}{2}: \qquad   &  \mathcal{C}_1(p)=\left\{\mathcal{L}_{[(\lambda,0;0)]}|\lambda\in\mathbb{N};\,\lambda\leq \tfrac{p}{2}-1\right\} \nonumber \\[2pt]
m=1: \qquad &  \mathcal{C}_2(p)=\left\{\mathcal{L}_{[(\lambda,\mu;0)]}|\lambda\in\mathbb{N};\, \lambda\leq \tfrac{p}{2}-1;\, \mu\in\{0,1\}\right\} \label{Cp} \\[2pt]
m=0: \qquad &    \mathcal{C}_3(p)=\left\{\mathcal{L}_{[(\lambda,\mu;0)]}|\lambda\in \tfrac{1}{2} \mathbb{N};\,\lambda\leq \tfrac{p}{2}-1;\, \mu\in\{0,\tfrac{1}{2},1\};\, \mu+\lambda\in \mathbb{N}\right\} \ , \nonumber
\end{align}
see Appendix~\ref{preserved set of defects} for the detailed analysis. We should also mention that the defects $\mathcal{L}_{[(\lambda,\frac{1}{2};0)]}$ which are in ${\cal C}_3(p)$ but not in ${\cal C}_2(p)$ simply anticommute (rather than commute) with the perturbation for $m=1$  --- which is the reason why they do not appear in ${\cal C}_2(p)$. 
Note that for $q=2$ the set of defects preserved by deformations with $m=0$ takes a slightly different form, as the defect $\mathcal{L}_{[(\frac{p-2}{4},\frac{1}{2};0)]}=\mathcal{L}_{(\frac{p-2}{4},\frac{1}{2};0)_+}+\mathcal{L}_{(\frac{p-2}{4},\frac{1}{2};0)_-}$ can be decomposed into the two simple defects $\mathcal{L}_{(\frac{p-2}{4},\frac{1}{2};0)_\pm}$. We therefore denote this set by $\overline{\mathcal{C}_3(p)}$.

In each case, one can show that in order for $\mathcal{C}_*(p)$ to be also present in the theory associated to $(p',q')$ we need $p'=p$ or $q'=p$; without loss of generality we may therefore take $p'=p$, and then $\mathcal{C}_*(p')$ has exactly the same structure as $\mathcal{C}_*(p)$. Thus we are interested in the flows 
\begin{align}
    \mathcal{SM}(p,q)\xrightarrow{\phi_{[(0,m;n)]}}\mathcal{SM}(p,q^\prime) \ .
\end{align}  
Note that because of the $c_{\text{eff}}$-theorem \cite{Castro-Alvaredo:2017udm}, we want $q>q^\prime$, and thus the case $q=2$ does not allow for any flows of this form since we need $q^\prime\geq 2$. 
On the other hand, $q'=2$ is interesting, and since $\mathcal{C}_3(p)$ in the UV and $\overline{\mathcal{C}_3(p)}$ in the IR are not isomorphic, it appears that $\mathcal{C}_3(p)$ cannot be preserved along such flows. However, there is a subalgebra in $\overline{\mathcal{C}_3(p)}$ that is isomorphic to $\mathcal{C}_3(p)$, where instead of the individual defects $\mathcal{L}_{(\frac{p-2}{4},\frac{1}{2};0)_\pm}$, we only consider their sum $\mathcal{L}_{[(\frac{p-2}{4},\frac{1}{2};0)]}=\mathcal{L}_{(\frac{p-2}{4},\frac{1}{2};0)_+}+\mathcal{L}_{(\frac{p-2}{4},\frac{1}{2};0)_-}$, see the discussion around \eqref{sum}. Thus $\mathcal{C}_3(p)$ can only flow to this subalgebra, and so in the following we will only analyse $\mathcal{L}_{[(\frac{p-2}{4},\frac{1}{2};0)]}$ and not $\mathcal{L}_{(\frac{p-2}{4},\frac{1}{2};0)_\pm}$ separately.

In order to determine the value of $q'$ that can appear, we determine next the quantum dimensions of these defect lines, 
\begin{align}
    d_{[(\lambda,\mu;0)]}(p,q)&=\frac{\sin\bigl(\pi (2\lambda+1)\frac{|q-p|}{2p}\bigr)\, \sin\bigl(\frac{\pi(2\mu+1)}{4}\bigr)}{\sin\bigl(\pi\frac{|q-p|}{2p}\bigr)\, \sin\bigl(\frac{\pi}{4}\bigr)} \ .\label{UV quantum dim}
\end{align}
We thus need to find the values of $q'$ for which $d_{[(\lambda,\mu;0)]}(p,q) = d_{[(\lambda,\mu;0)]}(p,q')$, where the defects are taken from ${\cal C}_*(p)$, see eq.~(\ref{Cp}),  depending on the form of the perturbing field. We can rewrite these expressions as 
\begin{align}
    \frac{\sin(\pi (2\lambda+1)\frac{|q-p|}{2p})}{\sin(\pi\frac{|q-p|}{2p})}
    &=\sum_{t=0}^{\lfloor\lambda\rfloor}(-1)^t\binom{2\lambda+1}{2s+1}
    \cos^{2(\lambda-t)}\left(\pi\frac{|q-p|}{2p}\right)\sin^{2t}\left(\pi\frac{|q-p|}{2p}\right)\nonumber\\
    &=\sum_{t=0}^{\lfloor\lambda\rfloor}(-1)^{3t+1}\binom{2\lambda+1}{2t+1}
    \sin^{2(\lambda-t)}\left(\pi\frac{q}{2p}\right)\cos^{2t}\left(\pi\frac{q}{2p}\right)\ . 
    \label{qdim as funcion of q}
\end{align}
The structure now depends on whether $\lambda\in\mathbb{N}$ (as is the case for ${\cal C}_1(p)$ and ${\cal C}_2(p)$, i.e.\ for $m\neq 0$), or whether $\lambda$ can also be half-integer (as is the case for $m=0$, i.e.\ ${\cal C}_3(p)$). 
\smallskip

\noindent {\bf Case 1:} If $\lambda\in\mathbb{N}$, then only even powers of the sine and cosine function appear, and the ratio is invariant under reflecting $q$ around a multiple of $p$, i.e.\ under the transformation $q \equiv sp + I \mapsto q' = sp -I$, where $I$ and $s$ are integers.\smallskip

\noindent {\bf Case 2:} If $\lambda$ is half-integer (as will be the case provided that $m=0$), we get odd powers of the sine and cosine function, and the ratio is invariant under reflecting $q$ around an odd multiple of $p$, i.e.\ under the transformation $q \equiv sp + I \mapsto q' = sp -I$, where $I$ is an integer, and $s$ is an odd integer.
\smallskip

\noindent Thus the topological defects predict flows of the form 
\be\label{mainresult}
\mathcal{SM}(p,sp+I) \ \to \ \mathcal{SM}(p,sp-I)  \ , 
\ee
where $I$ is always an integer, but $s$ is either\footnote{It seems plausible to assume that for $m=1$, $s$ is actually even, since otherwise more defect line dimensions would be preserved than required.}  
\begin{align}
\hbox{perturbation by $\phi_{[(0,m;n)]}$ with $m=0$:} \qquad & \hbox{$s$ is an odd integer} \label{m=0 means s odd}\\
\hbox{perturbation by $\phi_{[(0,m;n)]}$ with $m=1$:} \qquad & \hbox{$s$ is any integer.}
\end{align}
We should note that these flows are automatically compatible with the constraints that were discussed below eq.~(\ref{4.8}). In particular, since $q-q'=2I$ is an even integer, the parity of $q'$ is the same as that of $q$. In order to see that $u-u'=I$ is in fact even for $p$ and $q$ odd and $\phi_{[(0,m;n)]}$ a boson, we note that because of eqs.~\eqref{mF} and \eqref{worldsheet fermion number for u even} the deformation is bosonic if $m=0$. Then $s$ is an odd integer because of \eqref{m=0 means s odd}, and thus $I=q-sp$ is even. 

We should also mention that it is now essentially trivial to check that the preserved defect algebras are isomorphic between the UV and the IR. Indeed, since the relevant defect lines only depend on $p$, see eq.~(\ref{Cp}), the corresponding sets remain the same. Furthermore, since the defect lines all have $\nu=0$, their fusion rules are also essentially independent of $q$, and given that their quantum dimensions agree, they must form the same algebra in the IR as in the UV. While this analysis shows how the preserved defects get transformed under the RG flow, it would also be interesting to understand what happens to the other defects of the UV theory, and it may be possible to study this using the techniques of \cite{Fukusumi:2025fvb}.

Note that the only reason why the flows with $s$ an even integer seem to be forbidden for $m=0$, is that the quantum dimension of the defect lines with $\lambda$ half-integer (but not integer) pick up a minus sign under the flow. However, the product structure of the set of topological defect lines ${\cal C}_3(p)$ remains the same.\footnote{The reason for this is that there are two $1$-dimensional representations of the corresponding fusion ring that differ precisely by this sign.}  
These signs may also be related to the fact that the defects with $\lambda$ half-integer do not commute with the perturbation $\phi_{[(0,1:s)]}$, but rather anti-commute, i.e.\ they satisfy (\ref{preservation}) except for a relative sign between the two sides. (This is the reason why these defects are not elements of ${\cal C}_2(p)$, see also the comment below eq.~(\ref{Cp}).)

\subsubsection{Identifying the perturbing field}

Given the analysis of Section~\ref{sec:relevant}, it is natural to believe that the relevant deformation $\phi_{[(0,m;n)]}$ inducing the flow $\mathcal{SM}(p,sp+I)\to\mathcal{SM}(p,sp-I)$ is $\phi_{[(0,0;s)]}$ if $s$ is odd, and $\phi_{[(0,1;s)]}$ if $s$ is even
\begin{align}
    \mathcal{SM}(p,sp + I ) & \xrightarrow{\phi_{[(0,0;s)]}}\mathcal{SM}(p,sp - I ) \qquad (\hbox{$s$ odd}) \\
     \mathcal{SM}(p,sp + I )& \xrightarrow{\phi_{[(0,1;s)]}}\mathcal{SM}(p,sp - I ) \qquad (\hbox{$s$ even})  \ ,
\end{align} 
whose conformal dimension in either case is given by 
\be
\left. h([(0,0;s)]) \right|_{s \, {\rm odd}} = \left. h([(0,1;s)]) \right|_{s \, {\rm even}}  = 
\frac{ 2 sp - I(s-1)}{2 (sp + I)} \ .
\ee
This is the least relevant relevant field for $s=1$ and $I<\frac{p}{2}$, which includes in particular the familiar unitary flows. It also has the property that it is a $G$-descendant, see eq.~(\ref{mu value for susy primary}), as one would expect for a supersymmetry preserving flow. In general, however, a fine-tuned linear combination of different perturbing fields of the form $\phi_{[(0,m;n)]}$ may be required, so it is difficult to be very definite here.

\subsubsection{Checking the spin content of the defect Hilbert space}

Finally, we can also check whether the spin content of the defect Hilbert space of these defects is indeed RG invariant. For the topological defect line corresponding to $\mathcal{L} \equiv \mathcal{L}_{[(\lambda,\mu;0)]}$, the spin content of the twisted Hilbert space is 
\be
{\rm Spin}_{\cal L} = \left\{ h_{[(l_1,m_1;n_1)]} - h_{[(l_2,m_2;n_2)]}  \, \, \hbox{mod} \, \mathbb{Z}: {\cal N}_{[(\lambda,\mu;0)], [l_1,m_1;n_1)]}{}^{[l_2,m_2;n_2)]} \neq 0 \right\} \ . 
\ee
For a general defect, this is quite complicated to calculate, but it turns out that if all of $\mathcal{C}_3(p)$ is preserved, it is enough to determine the spin content for $\mathcal{L}_{[(\frac{1}{2},\frac{1}{2};0)]}$ as it gives the strongest restrictions on the existence of the RG flows --- the reason for this is that it generates all the other defects that are preserved.
Similarly, if only $\mathcal{C}_2(p)$ is preserved, it is enough to calculate the spin content of $\mathcal{L}_{[(1,0;0)]}$ and $\mathcal{L}_{[(0,1;0)]}$.
In any case, we have found experimentally that the spin content agrees if and only if the quantum dimensions agree,\footnote{We explain some details of the spin content analysis in Appendix~\ref{spin content as RG invariant}.} so this analysis does not give any new constraints. This is also what happened in the analysis of \cite{Nakayama:2024msv}, and it would be interesting to understand more conceptually why this seems to be the case.

\subsection{Examples}

In this section we exhibit a few examples of the various flows.  First of all, our proposed flows contain the familiar flows $\mathcal{SM}(m,m+2)\to\mathcal{SM}(m,m-2)$ between unitary models \cite{Poghossian:1987ngr,Kastor:1988ef,Poghosyan:2014jia}, which have been checked to leading order. In the following we shall therefore concentrate on flows that are not of this type. 

\subsubsection{$\mathcal{SM}(4,4s+I)\to\mathcal{SM}(4,4s-I)$}

For $p=4$ the allowed values of $q$ are of the form $q=2+4k$, where $k\in\mathbb{N}$. The set of 
defects that are preserved by $\phi_{[(0,0;s)]}$ for $s$ odd are 
\be
\mathcal{C}_3(4)=\{\mathcal{L}_{[(0,0;0)]},\mathcal{L}_{[(0,1;0)]},\mathcal{L}_{[(\frac{1}{2},\frac{1}{2};0)]},\mathcal{L}_{[(1,0;0)]},\mathcal{L}_{[(1,1;0)]}\} \ , \qquad (\hbox{$s$ odd})
\ee
while for $s$ even we consider instead the perturbation by $\phi_{[(0,1;s)]}$
\be
\mathcal{C}_2(4)=\{\mathcal{L}_{[(0,0;0)]},\mathcal{L}_{[(0,1;0)]},\mathcal{L}_{[(1,0;0)]},\mathcal{L}_{[(1,1;0)]}\}\ , \qquad (\hbox{$s$ even).}
\ee
The defects $\mathcal{L}_{[(0,0;0)]}$ and $\mathcal{L}_{[(0,1;0)]}$ have trivial quantum dimension (equal to $+1$), and therefore do not lead to interesting constraints. On the other hand, the defect lines $\mathcal{L}_{[(\frac{1}{2},\frac{1}{2};0)]}$ and $\mathcal{L}_{[(1,0;0)]}$  satisfy the algebra 
\begin{align}
    \begin{split}
        &\mathcal{L}_{[(\frac{1}{2},\frac{1}{2};0)]}\times\mathcal{L}_{[(\frac{1}{2},\frac{1}{2};0)]}=\mathcal{L}_{[(0,0;0)]}+\mathcal{L}_{[(1,0;0)]}+\mathcal{L}_{[(0,1;0)]}+\mathcal{L}_{[(1,1;0)]}\ ,\\
        &\mathcal{L}_{[(1,0;0)]}\times \mathcal{L}_{[(1,0;0)]} = \mathcal{L}_{[(0,0;0)]}\ ,\\
        &\mathcal{L}_{[(\frac{1}{2},\frac{1}{2};0)]}\times\mathcal{L}_{[(1,0;0)]}=\mathcal{L}_{[(\frac{1}{2},\frac{1}{2};0)]} \ .
    \end{split}\label{fusion algebra C_3(4)}
\end{align}
Note that for $q=2$ we have $\mathcal{L}_{[(0,0;0)]}=\mathcal{L}_{[(1,1;0)]}$ and $\mathcal{L}_{[(1,0;0)]}=\mathcal{L}_{[(0,1;0)]}$ and hence multiplicities of $2$ appear in \eqref{fusion algebra C_3(4)}. 

For odd $s$, the defect $\mathcal{L}_{[(\frac{1}{2},\frac{1}{2};0)]}$ generates the whole algebra,\footnote{For $q=2$, we mean by this defect the sum $\mathcal{L}_{[(\frac{1}{2},\frac{1}{2};0)]}=\mathcal{L}_{(\frac{1}{2},\frac{1}{2};0)_+}+\mathcal{L}_{(\frac{1}{2},\frac{1}{2};0)_-}$. \label{foot12}} and its quantum dimension and defect Hilbert space spin content are given (for small values of $q$) by 
\begin{table}[h]
\centering
\begin{tabular}{ |c|c|c|c|c|c|c|c| } 
 \hline
 $q$ & 2 & 6 & 10 & 14 & 18 & 22 & 26     \\ 
 \hline
 $d_{[(\frac{1}{2},\frac{1}{2};0)]}$ & 2 & 2 & -2 & -2 & 2 & 2 & -2    \\ 
 \hline
 $\text{Spin}_{[(\frac{1}{2},\frac{1}{2};0)]}$&$0;\pm\frac{1}{8}$& $0;\pm\frac{1}{8}$ &$0;\pm\frac{1}{8};\pm\frac{1}{4}$  & $0;\pm\frac{1}{8};\pm\frac{1}{4}$ & $0;\pm\frac{1}{8}$  &$0;\pm\frac{1}{8}$&$0;\pm\frac{1}{8};\pm\frac{1}{4}$  \\
 \hline
\end{tabular}
\caption{RG invariants of the defect line $\mathcal{L}_{[(\frac{1}{2},\frac{1}{2};0)]}$ in the $\mathcal{M}(4,q)$ series.}
\label{N=1 minimal models RG invariants for p=4, L1/2,1/2;0}
\end{table}

\noindent We notice the reflection symmetry around odd multiples of $4$ as expected. Note that the last flow, $\mathcal{SM}(4,6)\to\mathcal{SM}(4,2)$, is between unitary minimal models and therefore has been independently confirmed. From the perspective of this paper it is only consistent because the simple defect $\mathcal{L}_{[(\frac{1}{2},\frac{1}{2};0)]}$ of $\mathcal{SM}(4,6)$ flows to 
\be
\mathcal{L}_{[(\frac{1}{2},\frac{1}{2};0)]} [\mathcal{SM}(4,6)] \mapsto 
\mathcal{L}_{(\frac{1}{2},\frac{1}{2};0)_+}[\mathcal{SM}(4,2)]  
+\mathcal{L}_{(\frac{1}{2},\frac{1}{2};0)_-}[\mathcal{SM}(4,2)]  \ , 
\ee
see Footnote~\ref{foot12}. Interestingly, this seems to provide a counterexample to the statement that simple defects flow to simple defects under RG flows \cite{kikuchi2022}.
\smallskip

For even $s$, the defect $\mathcal{L}_{[(\frac{1}{2},\frac{1}{2};0)]}$ is not preserved, and we need to consider instead the other non-trivial defects, namely $\mathcal{L}_{[(1,0;0)]}$ and $\mathcal{L}_{[(1,1;0)]}$. Their quantum dimensions and defect Hilbert spin content however are simply equal to 
\be
d_{[(1,0;0)]} = d_{[(1,1;0)]} = 1 \ , \qquad \text{Spin}_{[(1,0;0)]} = \text{Spin}_{[(1,1;0)]}  = 0 \ , \qquad (q = 2 + 4k) \ , 
\ee
for all $q = 2 + 4k$, and hence we do not find any constraints from them. Note that this is compatible with the fact that any pair $q_1 = 2 + 4k_1$ and $q_2 = 2 + 4 k_2$ are connected to one another by a sequence of flows of the form $\mathcal{SM}(4,4s+I)\to\mathcal{SM}(4,4s-I)$ with $s$ integer, and since for both $s$ even and $s$ odd, these defect lines are preserved, all their quantum dimensions need to agree.

\subsubsection{$\mathcal{SM}(5,5s+I)\to\mathcal{SM}(5,5s-I)$}

As a second example let us consider the case where $p=5$. Now the different sets of defects 
$\mathcal{C}_*(5)$ take the form 
\be
\begin{array}{ll}
{\displaystyle \mathcal{C}_2(5) =   \{\mathcal{L}_{[(0,0;0)]},\mathcal{L}_{[(0,1;0)]},\mathcal{L}_{[(1,0;0)]},\mathcal{L}_{[(1,1;0)]}\}} \qquad & \hbox{($s$ even),} \\
{\displaystyle \mathcal{C}_3(5)=\{\mathcal{L}_{[(0,0;0)]},\mathcal{L}_{[(0,1;0)]},\mathcal{L}_{[(\frac{1}{2},\frac{1}{2};0)]},\mathcal{L}_{[(1,0;0)]},\mathcal{L}_{[(1,1;0)]},\mathcal{L}_{[(\frac{3}{2},\frac{1}{2};0)]}\}} \qquad
& \hbox{($s$ odd).}
\end{array}
\ee
Again, the defects $\mathcal{L}_{[(0,0;0)]}$ and $\mathcal{L}_{[(0,1;0)]}$ have trivial quantum dimension (and defect Hilbert space spin), and we shall therefore concentrate on the other defect lines, in particular $\mathcal{L}_{[(\frac{1}{2},\frac{1}{2};0)]}$, $\mathcal{L}_{[(1,0;0)]}$, and $\mathcal{L}_{[(\frac{3}{2},\frac{1}{2};0)]}$. They satisfy the fusion algebra 
\begin{align}
    \begin{split}
        &\mathcal{L}_{[(\frac{1}{2},\frac{1}{2};0)]}\times\mathcal{L}_{[(\frac{1}{2},\frac{1}{2};0)]}=\mathcal{L}_{[(0,0;0)]}+\mathcal{L}_{[(1,0;0)]}+\mathcal{L}_{[(0,1;0)]}+\mathcal{L}_{[(1,1;0)]}\ , \\
        &\mathcal{L}_{[(1,0;0)]}\times \mathcal{L}_{[(1,0;0)]} = \mathcal{L}_{[(0,0;0)]}\ , \\
        &\mathcal{L}_{[(\frac{1}{2},\frac{1}{2};0)]}\times\mathcal{L}_{[(\frac{1}{2},\frac{1}{2};0)]}=\mathcal{L}_{[(0,0;0)]}+\mathcal{L}_{[(1,0;0)]}\ , \\
        &\mathcal{L}_{[(\frac{1}{2},\frac{1}{2};0)]}\times\mathcal{L}_{[(1,0;0)]}=\mathcal{L}_{[(\frac{1}{2},\frac{1}{2};0)]}+\mathcal{L}_{[(\frac{3}{2},\frac{1}{2};0)]}\ , \\
        &\mathcal{L}_{[(\frac{1}{2},\frac{1}{2};0)]}\times\mathcal{L}_{[(\frac{3}{2},\frac{1}{2};0)]}=\mathcal{L}_{[(0,1;0)]}+\mathcal{L}_{[(1,1;0)]}\ , \\
        &\mathcal{L}_{[(1,0;0)]}\times\mathcal{L}_{[(\frac{3}{2},\frac{1}{2};0)]}=\mathcal{L}_{[(\frac{1}{2},\frac{1}{2};0)]} \ .
    \end{split}\label{fusion algebra C_3(5)}
\end{align}
Their quantum dimensions and defect Hilbert space spin contents are described in Table~\ref{N=1 minimal models RG invariants for p=5, L1/2,1/2;0}.
\begin{table}[h!]
\centering
\resizebox{15.0cm}{!}{
\begin{tabular}{ |c|c|c|c|c|c| } 
 \hline
 $q$ & 3 & 7 & 9 & 11 & 13 \\
  \hline
 $d_{[(\frac{1}{2},\frac{1}{2};0)]}$ & $\sqrt{3+\sqrt{5}}$ & $\sqrt{3+\sqrt{5}}$ & $\sqrt{3-\sqrt{5}}$ & $-\sqrt{3-\sqrt{5}}$ & $-\sqrt{3+\sqrt{5}}$ \\ 
 \hline
 $\text{Spin}_{[(\frac{1}{2},\frac{1}{2};0)]}$&$\pm\frac{3}{80};\pm\frac{1}{16};\pm\frac{27}{80}$& $\pm\frac{3}{80};\pm\frac{1}{16};\pm\frac{27}{80}$ &$\pm\frac{1}{80};\pm\frac{9}{80};\pm\frac{5}{16}$  & $\pm\frac{11}{80};\pm\frac{19}{80};\frac{1}{16}$ 
& $\pm\frac{7}{80};\pm\frac{17}{80};\pm\frac{3}{16}$  \\
\hline
\end{tabular}}
\vspace{0.5cm}

\resizebox{15.0cm}{!}{
\begin{tabular}{ |c|c|c|c|c|c|c|c|c|c| } 
 \hline
 $q$ & 3 & 7 & 9 & 11 & 13 & 17 & 19 \\ 
 \hline
 $d_{[(1,0;0)]}$ & $\sqrt{\frac{3+\sqrt{5}}{2}}$ & $\sqrt{\frac{3+\sqrt{5}}{2}}$ & $-\sqrt{\frac{3-\sqrt{5}}{2}}$ & $-\sqrt{\frac{3-\sqrt{5}}{2}}$ & $\sqrt{\frac{3+\sqrt{5}}{2}}$ & $\sqrt{\frac{3+\sqrt{5}}{2}}$ 
 $-\sqrt{\frac{3-\sqrt{5}}{2}}$  & $-\sqrt{\frac{3-\sqrt{5}}{2}}$ \\ 
 \hline
 $\text{Spin}_{[(1,0;0)]}$&$0;\pm\frac{2}{5}$& $0;\pm\frac{2}{5}$ &$0;\pm\frac{1}{5}$  & $0;\pm\frac{1}{5}$ & $0;\pm\frac{2}{5}$  & $0;\pm\frac{2}{5}$ & $0;\pm\frac{1}{5}$ \\ 
 \hline
\end{tabular}}
\vspace{0.5cm}

\begin{tabular}{ |c|c|c|c|c|c|c|c|c| } 
 \hline
 $q$ & 3 & 7 & 9 & 11 & 13 & 17 & 19 & 21 \\ 
 \hline
 $d_{[(\frac{3}{2},\frac{1}{2};0)]}$ & $\sqrt{2}$ & $\sqrt{2}$ & $-\sqrt{2}$ & $\sqrt{2}$ & $-\sqrt{2}$ & $-\sqrt{2}$ & $\sqrt{2}$ & $-\sqrt{2}$ \\ 
 \hline
 $\text{Spin}_{[(\frac{3}{2},\frac{1}{2};0)]}$&$\pm\frac{1}{16}$&$\pm\frac{1}{16}$&$\pm\frac{3}{16}$&$\pm\frac{1}{16}$&$\pm\frac{3}{16}$&$\pm\frac{3}{16}$&$\pm\frac{1}{16}$&$\pm\frac{3}{16}$ \\  
 \hline
\end{tabular}
\caption{RG invariants of the defect lines $\mathcal{L}_{[(\frac{1}{2},\frac{1}{2};0)]}$, $\mathcal{L}_{[(1,0;0)]}$, and $\mathcal{L}_{[(\frac{3}{2},\frac{1}{2};0)]}$ in the $\mathcal{M}(5,q)$ series for the first few values of $q$.}
\label{N=1 minimal models RG invariants for p=5, L1/2,1/2;0}
\end{table}

Let us first concentrate on the RG invariants of the defect lines  $\mathcal{L}_{[(\frac{1}{2},\frac{1}{2};0)]}$ and $\mathcal{L}_{[(\frac{3}{2},\frac{1}{2};0)]}$. They are preserved provided that $s$ is odd, and hence must exhibit a reflection symmetry around odd multiples of $5$ --- which they indeed do. 

On the other hand, the defect line $\mathcal{L}_{[(1,0;0)]}$ is perserved both for $s$ odd and $s$ even, and hence its RG invariants must exhibit a reflection symmetry around any multiple of $5$ --- which is also true.

\section{The supersymmetric setup}
\label{Supersymmetric defects}

Up to now we have considered the coset theory with spectrum (\ref{cosettheory}), but this theory is actually not superconformal since it arises from the ${\cal N}=1$ minimal model representations by imposing the GSO projection, see eq.~(\ref{GSO}). In this section we want to study the superconformal field theory instead. 

The `correct' definition of its spectrum is, however, a little bit subtle. If the spectrum is to contain the supercurrents (as it should, if the theory is `superconformal'), then its partition function {\bf cannot} be invariant under the $T$-transformation, $\tau\mapsto \tau+1$, but only under $\tau \mapsto \tau+2$. On the other hand, we would still expect the theory to be invariant under the $S$-modular transformation. Thus we shall take the spectrum to be the complete NS-NS spectrum, but without any R-R states.\footnote{From the perspective of the superconformal algebra, the Ramond sector is the $\mathbb{Z}_2$-twisted sector associated to $(-1)^{F}$.} More specifically, the complete set of NS representations of the ${\cal N}=1$ superconformal algebra are labelled by $[(l;n)]$, where $l+n\in\mathbb{Z}$ and we set 
\be
{\cal H}_{[(l;n)]} = {\cal H}_{[(l,0;n)]} \oplus {\cal H}_{[(l,1;n)]}  \ .
\ee
Then the spectrum of the ${\cal N}=1$ superconformal theory is 
\be\label{susytheory}
{\cal H}^{({\rm susy})} = \bigoplus_{[(l;n)]} \, {\cal H}_{[(l;n)]}  \otimes \tilde{{\cal H}}_{[(l;n)]}  = 
\bigoplus_{[(l;n)]}  \Bigl(  {\cal H}_{[(l,0;n)]} \oplus {\cal H}_{[(l,1;n)]} \Bigr) \otimes 
\Bigl(  \tilde{\cal H}_{[(l,0;n)]} \oplus \tilde{\cal H}_{[(l,1;n)]} \Bigr) \ ,
\ee
i.e.\ with respect to the coset algebra, it is of extension type. Note that the modular $S$-matrix closes on these representations, and one finds from (\ref{coset S matrix}) that 
\be
\begin{split}
   \mathcal{S}^{({\rm susy})}_{(\lambda;\nu),(\lambda^\prime;\nu^\prime)}= \frac{4}{\sqrt{pq}}\, \sin\left(\pi(2\lambda+1)(2\lambda^\prime+1)\frac{|q-p|}{2p}\right) \sin\left(\pi(2\nu+1)(2\nu^\prime+1)\frac{|q-p|}{2q}\right)\! . 
\end{split}\label{N=1S matrix}
\ee
Since the resulting theory is now diagonal with respect to the extended characters, essentially the same analysis as for the bosonic coset theory above applies. In particular, the defects of the superconformal theory are labelled by the NS-representations $[(\lambda;\nu)]$ with $\lambda+\nu\in\mathbb{Z}$, and they act on the NS-sector states via the analogue of (\ref{actionform}), i.e.\ as 
\begin{align}\label{actionforms}
    \mathcal{L}_{[(\lambda;\nu)]}|\phi_{(l;n)}\rangle =\frac{{\cal S}^{({\rm susy})}_{(\lambda;\nu)(l;n)}}{{\cal S}^{({\rm susy})}_{(0;0)(l;n)}}\, |\phi_{(l;n)}\rangle \ .
\end{align}
Another way of saying this is that the extended symmetry includes the supercurrent modes $G_{r}$  that intertwine between the two NS-representations $[(l,0;n)]$ and $[(l,1;n)]$. These two representations have the same $\mathcal{S}$-matrix elements with any NS-sector representation, and thus the two defects $\mathcal{L}_{[(\lambda,0;\nu)]}$ and $\mathcal{L}_{[(\lambda,1;\nu)]}$ act identically, namely as $\mathcal{L}_{[(\lambda;\nu)]}$.

Obviously, there is now no analogue of the coset defect corresponding to $(-1)^{F_s}$, see eq.~(\ref{010act}). Furthermore, the chiral fermion number defect of Section~\ref{sec:worldsheetferm} cannot be realised by such a defect since it does not commute with the entire extended chiral algebra. (It anti-commutes with the supercurrent modes.)

The natural analogue of the perturbation in Section~\ref{sec:deformations} is now a perturbation by $G_{-\frac{1}{2}}\phi_{[(0;n)]}$, where $n$ is integer. This perturbation now preserves the defects (in the sense of eq.~(\ref{preservation})) that lie in the set, cf.\ eq.~(\ref{Cp})
\be
 \mathcal{C}^{({\rm susy})}_2(p)=\left\{\mathcal{L}_{[(\lambda;0)]}|\lambda\in\mathbb{N};\, \lambda\leq \tfrac{p}{2}-1 \right\} \label{Cps} \ .
 \ee
 The analysis around eq.~(\ref{qdim as funcion of q}) goes through essentially unchanged, and we conclude that\footnote{In the present set-up only Case 1 is relevant.} the defect considerations suggest the flows 
\be\label{susyRG}
\mathcal{SM}(p,sp + I )  \xrightarrow{\phi_{(0;s)}}\mathcal{SM}(p,sp - I ) \ ,
\ee
where $s$ and $I$ are arbitrary integers.

\section{Conclusions}\label{sec:concl}

In this paper we have studied the RG flows of the ${\cal N}=1$ superconformal models, using similar techniques to what was done before for the bosonic minimal models in \cite{Nakayama:2024msv}. In particular, we have identified the subalgebra of those topological defects that commute with the perturbation, and then checked whether they are in fact also present in the IR theory. Our analysis was first done using the coset description of the bosonic subalgebra for which the fixed-point at the center of the Kac table had to be resolved, see the discussion in Section~\ref{sec:2.3}. We have subsequently also looked at the actual superconformal theory that, from the perspective of the bosonic coset description, is of modular extension type. In either case, we could find RG flows, see eqs.~(\ref{mainresult}) and (\ref{susyRG}) that respect the topological defects and have a similar structure to what was proposed in \cite{Nakayama:2024msv}.

Some of the complications of our analysis had to do with the fact that we are considering a superconformal (rather than bosonic) field theory, see e.g.\ the discussion in Section~\ref{sec:symmetries}, and it should therefore be possible to generalise our discussion to other superconformal models, in particular to the ${\cal N}=2$ minimal models \cite{Boucher:1986bh,DiVecchia:1986fwg}, or theories with even more supersymmetry. However, there are interesting differences that make this generalisation non-trivial. In particular, the integrable RG flows of the ${\cal N}=2$ minimal models are infinite distance in the Zamolodchikov metric, and we have checked  that they do not preserve the quantum dimensions of the defect lines that commute with the perturbation. Thus, a naive generalisation of the ideas of this paper to ${\cal N}=2$ RG flows does not seem to work.\footnote{The ${\cal N}=2$ minimal models also differ from the Virasoro and ${\cal N}=1$ super Virasoro minimal models in that they are all necessarily unitary \cite{Eholzer:1996zi}.} However, it may be possible to analyse the constraints from these defects on exactly marginal deformations, and it would be interesting to explore this in more detail. This would, in particular, be relevant in the context of string theory since spacetime supersymmetry requires at least ${\cal N}=2$ superconformal symmetry on the worldsheet \cite{Banks:1987cy}.

\section*{Acknowledgements}

This paper is largely based on the Master thesis \cite{Merkens} of one of us (LM). MRG thanks 
Yu Nakayama for useful conversations, as well as Rajesh Gopakumar and Wolfgang Lerche for discussions about related topics. We also acknowledge useful correspondences with Yoshiki Fukusumi and Ken Kikuchi after a first version of the paper appeared on the arXiv. Finally, we thank the anonymous referee for various useful suggestions concerning the interplay between the defects and the fermion number symmetry. The work of the group at ETH is supported by a personal grant of MRG from the Swiss National Science Foundation, by the Simons Foundation grant 994306 (Simons Collaboration on Confinement and QCD Strings), as well as the NCCR SwissMAP that is also funded by the Swiss National Science Foundation.

\appendix
\label{appendix}

\section{Cosets for the non-unitary ${\cal N}=1$ minimal models} \label{su2frac}

In this appendix we give a brief review of the coset construction for the non-unitary ${\cal N}=1$ minimal models.

\subsection{The fractional level $\mathfrak{su}(2)$ theories}

Let us begin by summarising the salient features of the fractional $\mathfrak{su}(2)$ theories. While the Wess-Zumino action is not well-defined for fractional levels, one can still define the current algebra generators via 
\begin{align}
[J^3_m,J^3_n] = \frac{k}{2} m \delta_{m,-n} \ , \quad [J^3_m,J^\pm_n] =  \pm J^\pm_{m+n} \ , \quad 
[J^+_m,J^-_n] = 2 J^3_{m+n} + k m \delta_{m,-n} \ ,
\end{align}
and the theory is still conformal because also the definition of the Sugawara stress-energy tensor still makes sense (as long as $k\neq -2$). However, the resulting theories are typically not rational, i.e.\ they possess a priori infinitely many representations, and their characters do not transform nicely among themselves.

However, there are special levels, the so-called {\bf admissible} levels \cite{Kac:1988qc}, for which one can identify a finite set of highest weight representations whose characters, at least formally, transform into one another. The admissible levels are of the form 
\begin{align}
    k=\frac{t}{u} \ ,
\end{align}
where
\begin{align}
    u\in\mathbb{N}\ , && t\in\mathbb{Z} \textbackslash \{0\} \ ,  && \text{gcd}(t,u)=1,  && t=2u-2\geq0 \ . 
\end{align}
For the case at hand, we have 
\be
k=\frac{2\max(p,q)}{|q-p|}-2 \ , \qquad t = \min(p,q) \ , \ \ u = \frac{|q - p|}{2} \ .
\ee
The {\bf admissible} representations of ${\mathfrak{su}}(2)_{k}$ are then those highest weights $\lambda$ that can be decomposed into two integrable weights $\lambda^I$ and $\lambda^F$ as
\begin{align}\label{A.5}
\lambda = \lambda^I-(k+2)\lambda^F \ , 
\end{align}
where $\lambda^I$ and $\lambda^F$ are integrable weights at levels $k^I$ and $k^F$, respectively, with
\begin{align}
k^I=u(k+2)-2\geq0 \ , \qquad     \qquad k^F=u-1\geq 0 \label{k0kF} \ .
\end{align}
Then $\lambda$ is at level $k^I - (k+2) k^F = k$. The second inequality in (\ref{k0kF}) 
is simply a consequence of the fact that $|q-p|\geq 2$, whereas the first inequality is a consequence of the fact that $u(k+2) = \max(p,q)>2$. 

Since an admissible representation at level $k$ consists of two highest-weight representations at integer levels $k^I$ and $k^F$, there are $(k^I + 1)(k^F + 1)$ such representations for each $k$, and the conformal dimension of the corresponding highest-weight states equals
\begin{align}
    h_\lambda=\frac{\lambda(\lambda+1)}{k+2} \ .
\end{align}
By calculating the characters of the admissible representations, the modular $\mathcal{S}$ matrix can be determined as
\begin{align}
    \begin{split}
        \mathcal{S}_{\lambda,\lambda^\prime}=&\sqrt{\frac{2}{u^2(k+2)}}(-1)^{2\lambda^{\prime F}(2\lambda^I+1)+2\lambda^F(2\lambda^{\prime I}+1)}\\
        &\times e^{-i \pi 2\lambda^{\prime F}2\lambda^F(k+2)}\sin\left(\frac{\pi(2\lambda^I+1)(2\lambda^{\prime I}+1)}{k+2}\right) \ .
    \end{split}
    \label{fractional level WZW S matrix}
\end{align}
As previously mentioned, the fusion coefficients of the fractional-level WZW models obtained from (\ref{fractional level WZW S matrix}) and the Verlinde formula are sometimes negative. 
This will not be a problem for us, however, as we can restrict ourselves to the states for which the fractional weight $\lambda^F=0$, see Section~\ref{app:fractionalcoset} below.  In this case, the fusion coefficients simplify, and we obtain simply the familiar fusion rules of the $\lambda^I$ representations at level $k^I$, which are non-negative.

\subsection{The non-unitary cosets} \label{app:fractionalcoset}

With these preparations in hand, we can now explain the structure of the non-unitary coset theories
\begin{align}
    \frac{{\mathfrak{su}}(2)_k\oplus {\mathfrak{su}}(2)_2}{{\mathfrak{su}}(2)_{k+2}}\ , \qquad \hbox{with} \quad    k=\frac{2\max(p,q)}{|q-p|}-2 \ .
\end{align}
Solving for $p$ and $q$ the identification is simply 
\begin{align}
    p=k^I+2 && q=k^I+2u+2=(k+2)^I+2 \ .
\end{align}
The coset representations will again be labeled by $(\lambda,\mu;\nu)$, which are weights at level $k$, $2$ and $k+2$. The selection rule takes the form 
\begin{align}
    \lambda^I-(k+2)\lambda^F+\mu-\nu^I+(k+4)\nu^F\in \mathbb{Z}\ ,
\end{align}
which is equivalent to the two conditions
\begin{align}
        \lambda^F=\nu^F \ , \qquad         \lambda^I+\mu-\nu^I\in\mathbb{Z}\ . 
\end{align}
The field identifications are more complicated than in the unitary case, however. As for the familiar unitary coset models, we have the field identifications coming from the outer automorphism,
\begin{align}
    (\lambda,\mu;\nu)\mapsto \mathcal{T}(\lambda,\mu;\nu) \equiv (\tau_k\lambda,\tau_2\mu;\tau_{k+2}\nu)=\bigl(\tfrac{k}{2}-\lambda,1-\mu;\tfrac{k+2}{2}-\nu\bigr)\ ,
\end{align}
but now there is an additional identification often denoted by `charge conjugation' that identifies
\begin{align}
    (\lambda,\mu;\nu) \mapsto {\cal C}(\lambda,\mu;\nu)\equiv (c_k\lambda,c_2\mu;c_{k+2}\nu)\ .
\end{align}
Here $c_k$ acts on the fractional and integral weight $\lambda$ at level $k=\frac{t}{u}$ as
\begin{align}
    c_k\lambda^I=\frac{k^I}{2}-\lambda^I\ , \qquad     
    c_k\lambda^F=\frac{u}{2}-\lambda^F-u\delta_{\lambda^F,0}\ .
\end{align}
To obtain all field identifications, we must consider arbitrary combinations of $\mathcal{T}$ and ${\cal C}$. Since $\mathcal{T}^2 =  {\cal C}^2 = 1$ the relevant transformations can be thought of as words made from alternating letters ${\cal T}$ and ${\cal C}$ --- the two transformations do not commute with one another --- subject to the relation that $(\mathcal{T} {\cal C})^{u}=1$. As a consequence, generically $2u$ primary fields are identified in this manner. One can show that in every orbit of this chain there are exactly two coset fields with $\lambda^F = 0 = \nu^F$. 
Thus, we can always choose to represent the coset fields by the triplets $(\lambda^I, \mu; \nu^I)$ at integer levels $k^I$, $2$, and $(k+2)^I = k^I + 2u$, respectively. With this gauge choice, the only remaining identification is
\begin{align}
    (\lambda^I,\mu;\nu^I)\sim (\tau_{k^I}\lambda^I,\tau_2^u\mu;\tau_{k^I+2u}\nu^I)\ . \label{field identification1}
\end{align}
Note that for $u=\frac{|q-p|}{2}$ even the field identification does not change $\mu$, while for $u=\frac{|q-p|}{2}$ odd, it interchanges $\mu=0$ and $\mu=1$. 

\subsection{Fusion rules for even models}\label{Fusion rules for even models}

The fusion coefficients for the models with even $p$ and $q$ are essentially given by the same formula as for $p,q$ odd, see eq.~\eqref{fusionrules}, except for the fusions involving $(\frac{p-2}{4},\frac{1}{2};\frac{q-2}{4})_\pm$. The fusion involving these fields are 
\begin{align}
        &\left(\tfrac{p-2}{4},\tfrac{1}{2};\tfrac{q-2}{4}\right)_\pm\otimes\left(\tfrac{p-2}{4},\tfrac{1}{2};\tfrac{q-2}{4}\right)_\pm=\bigoplus_{\substack{(\lambda_3,0;\nu_3)\\\lambda_3,\nu_3\in 2\mathbb{Z}}}{\cal N}_{{\frac{p-2}{4}} {\frac{p-2}{4}}}^{[k^I]\,\, \lambda_3} \, 
            {\cal N}_{{\frac{q-2}{4}} {\frac{q-2}{4}}}^{[(k+2)^I]\,\, \nu_3} \, [(\lambda_3,0;\nu_3)] \\[3pt]
        &\left(\tfrac{p-2}{4},\tfrac{1}{2};\tfrac{q-2}{4}\right)_\mp\otimes\left(\tfrac{p-2}{4},\tfrac{1}{2};\tfrac{q-2}{4}\right)_\pm=\bigoplus_{\substack{(\lambda_3,0;\nu_3)\\\lambda_3,\nu_3\in 2\mathbb{Z}+1}}{\cal N}_{{\frac{p-2}{4}} {\frac{p-2}{4}}}^{[k^I]\,\, \lambda_3} \, 
            {\cal N}_{{\frac{q-2}{4}} {\frac{q-2}{4}}}^{[(k+2)^I]\,\, \nu_3} \, [(\lambda_3,0;\nu_3)]\\[3pt]
        &\left[\left(\lambda,\tfrac{1}{2};\nu\right)\right]\otimes\left(\tfrac{p-2}{4},\tfrac{1}{2};\tfrac{q-2}{4}\right)_\pm=\frac{1}{2} \bigoplus_{(\lambda_3,\mu_3;\nu_3)}{\cal N}_{\lambda \frac{p-2}{4}}^{[k^I]\,\, \lambda_3} \, {\cal N}_{\frac{1}{2}\, \frac{1}{2}}^{[2]\,\, \mu_3} \, 
            {\cal N}_{\nu \frac{q-2}{4}}^{[(k+2)^I]\,\, \nu_3} \, [(\lambda_3,\mu_3;\nu_3)]\nonumber \\[3pt]
        &\big[\left(\lambda,0;\nu\right)\big]_{\lambda,\nu\in2\mathbb{\mathbb{Z}}}\otimes\left(\tfrac{p-2}{4},\tfrac{1}{2};\tfrac{q-2}{4}\right)_\pm=\frac{1}{2}\bigoplus_{\substack{(\lambda_3,\frac{1}{2};\nu_3)
        \\ \neq(\frac{p-2}{4},\frac{1}{2},\frac{q-2}{4})}}{\cal N}_{\lambda \frac{p-2}{4}}^{[k^I]\,\, \lambda_3} \,  {\cal N}_{\nu \frac{q-2}{4}}^{[(k+2)^I]\,\, \nu_3} \, [(\lambda_3,\tfrac{1}{2};\nu_3)] \nonumber \\
            &\qquad\qquad\qquad\qquad\qquad\qquad\qquad\qquad\qquad\quad\bigoplus\left(\tfrac{p-2}{4},\tfrac{1}{2};\tfrac{q-2}{4}\right)_\pm\\[3pt]
        &\big[\left(\lambda,0;\nu\right)\big]_{\lambda,\nu\in2\mathbb{\mathbb{Z}}+1}\otimes\left(\tfrac{p-2}{4},\tfrac{1}{2};\tfrac{q-2}{4}\right)_\pm=\frac{1}{2}\bigoplus_{\substack{(\lambda_3,\frac{1}{2};\nu_3)\\\neq(\frac{p-2}{4},\tfrac{1}{2},\frac{q-2}{4})}}{\cal N}_{\lambda \frac{p-2}{4}}^{[k^I]\,\, \lambda_3} \, 
            {\cal N}_{\nu \frac{q-2}{4}}^{[(k+2)^I]\,\, \nu_3} \, [(\lambda_3,\tfrac{1}{2};\nu_3)] \nonumber \\ 
            &\qquad\qquad\qquad\qquad\qquad\qquad\qquad\qquad\qquad\quad\bigoplus\left(\tfrac{p-2}{4},\tfrac{1}{2};\tfrac{q-2}{4}\right)_\mp\\[3pt]
        &\big[\left(\lambda,0;\nu\right)\big]_{\lambda,\nu\in\mathbb{\mathbb{Z}}+\frac{1}{2}}\otimes\left(\tfrac{p-2}{4},\tfrac{1}{2};\tfrac{q-2}{4}\right)_\pm=\frac{1}{2}\bigoplus_{(\lambda_3,\frac{1}{2};\nu_3)}{\cal N}_{\lambda \frac{p-2}{4}}^{[k^I]\,\, \lambda_3} \, 
            {\cal N}_{\nu \frac{q-2}{4}}^{[(k+2)^I]\,\, \nu_3} \, [(\lambda_3,\tfrac{1}{2};\nu_3)] \ . 
\end{align}%
Note that for $p$ and $q$ even, $u$ is necessarily odd, see eq.~(\ref{udef}), and thus according to \eqref{field identification}, the field identification interchanges $\mu=0$ and $\mu=1$. As a consequence, it is sufficient to restrict ourselves to the cases $\mu=0$ and $\mu=\frac{1}{2}$.

\section{Defect perturbations}\label{app:defect}

In this appendix we collect some of the more technical material concerning the analysis of defect lines in the perturbed theory. 

\subsection{Defects preserved by $\phi_{[(0,m;n)]}$}\label{preserved set of defects}

The defect $\mathcal{L}_{[(\lambda,\mu;0)]}$ is preserved by the perturbation by $\phi_{[(0,m;n)]}$ provided that \linebreak eq.~(\ref{preservation}) holds. If $|\psi\rangle \in {\cal H}_{[(L,M;N)]}$, then the defects act as 
\be
\mathcal{L}_{[(\lambda,\mu;\nu)]} \, |\psi\rangle = \frac{\mathcal{S}_{(\lambda,\mu;0)(L,M;N)}}{\mathcal{S}_{(0,0;0)(L,M;N)}}\,  |\psi\rangle \ , 
\ee
and eq.~(\ref{preservation}) is simply the condition that 
\be
 \frac{\mathcal{S}_{(\lambda,\mu;0)(L,M;N)}}{\mathcal{S}_{(0,0;0)(L,M;N)}} =  \frac{\mathcal{S}_{(\lambda,\mu;0)(L,\hat{M};\hat{N})}}{\mathcal{S}_{(0,0;0)(L,M;N)}}\qquad
\hbox{for all}\ [(L,\hat{M};\hat{N})] \in [(0,m;n)] \otimes [(L,M;N)] \ . 
\ee
Because of the product structure of the $\mathcal{S}$ matrix of eq.~(\ref{coset S matrix}),\footnote{The modifications of the $S$-matrix due to the fixed-point resolution do not modify this result; we shall therefore assume that neither field is the fixed-point field.} only the factor coming from $\hat{\mathfrak{su}}(2)_2$ contributes, and the ratio simplifies to 
\be
 \frac{\mathcal{S}_{(\lambda,\mu;0)(L,M';N')}}{\mathcal{S}_{(0,0;0)(L,M;N)}} \Bigl( \frac{\mathcal{S}_{(\lambda,\mu;0)(L,M;N)}}{\mathcal{S}_{(0,0;0)(L,M;N)}} \Bigr)^{-1}    = \frac{\sin\bigl(\pi\frac{(2\mu+1)(2\hat{M}+1)}{4}\bigr)}{\sin \bigl(\pi\frac{(2\mu+1)(2M+1)}{4}\bigr)}\, \frac{\sin (\pi\frac{2M+1}{4})}{\sin(\pi\frac{2\hat{M}+1}{4})} \stackrel{!}{=} 1 \ . \label{condi}
 \ee
Furthermore, the possible values of $\hat{M}$ are constrained as
\begin{align}
m=0: \qquad &\hat{M} = M \\
m=\tfrac{1}{2}: \qquad & \hat{M} = \left\{ \begin{array}{cl} \frac{1}{2} \quad & \hbox{if $M=0$ or $M=1$} \\
\hbox{$0$ or $1$}  \quad & \hbox{if $M = \tfrac{1}{2}$} \end{array} \right. \\
m = 1: \qquad & \hat{M} = \left\{ \begin{array}{cl} 1 \quad & \hbox{if $M=0$ } \\
\tfrac{1}{2} \quad & \hbox{if $M=\tfrac{1}{2}$ } \\
0 \quad & \hbox{if $M =1$} \end{array} \right.
\end{align}
Thus, if $m=0$, (\ref{condi}) is automatically satisfied, and this leads to ${\cal C}_3(p)$. If $m=1$, on the other hand, the condition becomes 
\be
m=1: \qquad \sin \bigl(\tfrac{3 (2\mu +1)\pi}{4}\bigr)  \sin (\tfrac{\pi}{4}) \stackrel{!}{=} \sin \bigl(\tfrac{(2\mu +1)\pi}{4}\bigr)  \sin (\tfrac{3\pi}{4}) \ , 
\ee
which is true provided that $\mu=0,1$, but not for $\mu=\frac{1}{2}$, see ${\cal C}_2(p)$. [Note, though, that for $\mu=\frac{1}{2}$, the two sides only differ by a sign.] Finally, if $m=\frac{1}{2}$, the condition becomes 
\be
m=\tfrac{1}{2}: \qquad \sin \bigl(\tfrac{(2\mu +1)\pi}{4}\bigr)  \stackrel{!}{=} \sin \bigl(\tfrac{3 (2\mu +1)\pi}{4}\bigr) \stackrel{!}{=} \sin \bigl(\tfrac{(2\mu +1)\pi}{2}\bigr)  \sin (\tfrac{\pi}{4}) \ , 
\ee
which is only true for $\mu=0$, but not for $\mu=\frac{1}{2}$ or $\mu=1$, see  ${\cal C}_1(p)$. This completes the analysis. 

\subsection{Defect Hilbert space spin content as RG invariant}
\label{spin content as RG invariant}

In this appendix we calculate the spin content of the defect line $\mathcal{H}_{[(\frac{1}{2},\frac{1}{2};0)]}$. The left- and right-movers appearing in $\mathcal{H}_{[(\frac{1}{2},\frac{1}{2};0)]}$ are related to one another by fusion with $\phi_{[(\frac{1}{2},\frac{1}{2};0)]}$, and this maps\footnote{For the analysis of the spin content the modifications coming from the fixed-point resolution do not matter since this only affects the fusion rules minimally --- in particular, the fusion rules of the combinations (\ref{sum}) are always described by the generic fusion rules.}
\begin{align}
[(\tfrac{1}{2},\tfrac{1}{2};0)] \otimes [(m,0;k)] & = \bigoplus_{n=m\pm \frac{1}{2}}  [(n,\tfrac{1}{2};k)]  \label{first} \\
[(\tfrac{1}{2},\tfrac{1}{2};0)] \otimes [(m,\tfrac{1}{2};k)] & = \bigoplus_{n=m\pm \frac{1}{2}} \Bigl(  [(n,0;k)] \oplus  [(n,1;k)] \Bigr) \label{middle} \\
[(\tfrac{1}{2},\tfrac{1}{2};0)] \otimes [(m,1;k)] & = \bigoplus_{n=m\pm \frac{1}{2}}  [(n,\tfrac{1}{2};k)] \ . \label{third}
\end{align}
The two conformal dimensions that appear in (\ref{middle}) make up the NS-sector representation of the ${\cal N}=1$ superconformal field theory,\footnote{In the first and third line, the right-hand side is in the R-sector.} and their conformal dimension therefore differs by a half-integer. As a consequence we can directly work with the ${\cal N}=1$ representations, but work modulo $\frac{1}{2}\mathbb{Z}$. In this language the representation $[(\frac{1}{2},\frac{1}{2};0)]$ corresponds to $(r=2,s=1)$, see eq.~(\ref{translation}), and the conformal dimensions are given by eq.~(\ref{susy conformal dimensions}). Furthermore, the fusion rules simply become
\be
(r,s) \otimes (2,1) = (r+1,s) \oplus (r-1,s) \ . 
\ee
If the field on the left-hand side is in the R sector, i.e.\ $r+s$ is odd, then the difference in conformal dimension equals 
\begin{align}
    h_{r\pm 1,s}-h_{r,s}&= \frac{(q(r\pm 1)-sp)^2-(|q-p|)^2}{8pq} -\frac{1}{16}-\frac{(qr-sp)^2-(|q-p|)^2}{8pq}\\
    &=\frac{q^2\pm 2q^2r\mp 2pqs}{8pq} -\frac{1}{16} =\frac{q(1\pm 2r)}{8p} \mp \frac{s}{4} -\frac{1}{16} \ .
\end{align}
If $r+s$ is even, i.e.\ if the field on the left-hand side is in the NS sector and we combine (\ref{first}) and (\ref{third}),  the analysis is the same, except that the $\frac{1}{16}$ term appears with the opposite sign. Altogether we therefore learn that the difference in conformal dimension equals 
\be
\Delta_{r,s}(p,q) = \frac{q(1\pm 2r)}{8p} \mp \frac{s}{4} + (-1)^{r+s} \frac{1}{16} \ ,
\ee
where $r$ and $s$ take the values $r=1,\ldots,p-2$ and $s=1,\ldots q-1$. Modulo $\frac{1}{2}\mathbb{Z}$, $\Delta_{r,s}(p,q)$ is then equal to 
\begin{align}
\Delta_{r,s} (p,q)=     \frac{q(1\pm 2r)}{8p}+\begin{cases}
        \frac{-3}{16}& \text{for } r \text{ odd and } s \text{ odd}\\
        \frac{-1}{16}& \text{for } r \text{ odd and } s \text{ even}\\
        \frac{-5}{16}& \text{for } r \text{ even and } s \text{ odd}\\
        \frac{1}{16}& \text{for } r \text{ even and } s \text{ even}
    \end{cases}
\end{align}
For the behaviour of the spin content along the RG flow $\mathcal{SM}(p,q)\to\mathcal{SM}(p,q^\prime)$, the second term is irrelevant since it is independent of $p$ and $q$ (as long as $q\neq 2$). 

The conditions on $p$ and $q$ in (\ref{choice p q}) imply that $q=p+2u$ with $\text{gcd}\left(p,u\right)=1$, and thus we are interested in the spin content as a function of $u$, 
\be
\Delta'_{r,s} (u) = \frac{(1\pm 2r)}{8} + \frac{u (1\pm 2r)}{4 p}  \cong \frac{u}{2 p} \frac{(1\pm 2r)}{2}  \ . 
\ee
The spin spectrum thus stays the same provided that for each $r$, we can find an $r'$ such that 
\be
\frac{u}{p} \frac{(1\pm 2r)}{2} = \frac{u'}{p} \frac{(1\pm 2r')}{2}  \quad \mod\mathbb{Z} \ ,
\ee
where both $r$ and $r'$ take values in $r,r'\in\{1,\ldots, p-2\}$. This condition is satisfied provided that 
\be
u = u'  \quad \mod 2p\, \mathbb{Z} \ ,
\ee
and it is not difficult to see that this is also a necessary condition.
Thus, the condition that the defect space spin content remains the same yields the same restrictions on RG flows as the invariance of the quantum dimensions.

\end{document}